\documentclass[onecolumn,times,11pt,a4]{article}
\usepackage{amsmath}
\usepackage{amsthm}
\usepackage[latin2]{inputenc}
\usepackage{graphicx}
\usepackage{psfrag}
\usepackage{tikz}
\usepackage{tabularx}
\usepackage{comment}
\usetikzlibrary{shapes,arrows}
\RequirePackage{fullpage}

\usepackage[margin=0.96in]{geometry}    

\newtheorem{theorem}{Theorem}[section]
\newtheorem{lemma}[theorem]{Lemma}

\newtheorem{proposition}[theorem]{Proposition}

\newtheorem{defin}[theorem]{Definition}
\newtheorem{definition}[theorem]{Definition}
 
\DeclareMathAlphabet{\mathcal}{OMS}{cmsy}{m}{n}

\psfragscanon

\usepackage{tweaklist}
\renewcommand{\enumhook}{\setlength{\itemsep}{-0.5ex}} 
\renewcommand{\itemhook}{\setlength{\itemsep}{-0.5ex}}

\newcommand{\I}{\mathcal{I}}

\newcommand{\executeiffilenewer}[3]{%
\ifnum\pdfstrcmp{\pdffilemoddate{#1}}%
{\pdffilemoddate{#2}}>0%
{\immediate\write18{#3}}\fi%
} 

\newcommand{\svg}[2]{\def\svgwidth{#1}%
\executeiffilenewer{#2.svg}{#2.pdf}%
{inkscape -z -D --file=#2.svg %
--export-pdf=#2.pdf --export-latex}%
{\input{#2.pdf_tex}}}

\newcommand{\randset}{\textup{RandomSet}}
\newcommand{\derandset}{\textup{DeterministicSets}}
\newcommand{\mc}{\textsc{Vertex Multicut}}
\newcommand{\mcc}{\textsc{Multicut Compression}}
\newcommand{\mccs}{\textsc{Multicut Compression$^*$}}
\newcommand{\bmcc}{\textsc{Bipedal Multicut Compression$^*$}}
\newcommand{\redi}[1]{I/{#1}}
\newcommand{\torso}{\textup{\texttt{torso}}}
\newcommand{\ora}[1]{\overrightarrow{#1}}

\newcommand{\size}{p}
\newcommand{\pairs}{k}
\newcommand{\wleg}{\widehat W}

\begin{document}

\title{Fixed-parameter tractability of multicut parameterized
by the size of the cutset\footnote{A preliminary version of the paper was presented at STOC 2011~\cite{marx-razgon-multicut}.
Research of the second author was 
    supported by the European Research Council (ERC) grant
    ``PARAMTIGHT: Parameterized complexity and the search for tight
    complexity results,'' reference 280152 and grant OTKA NK105645.}}
\author{D\'aniel Marx\thanks{Institute for Computer Science and Control, Hungarian Academy of Sciences (MTA SZTAKI),
    \texttt{dmarx@cs.bme.hu}}
\and {Igor Razgon\thanks{Department of Computer Science and Information Systems, Birkbeck, University of
    London. \texttt{igor@dcs.bbk.ac.uk}}}}
\maketitle
\begin{abstract}
  Given an undirected graph $G$, a collection $\{(s_1,t_1), \dots,
  (s_{\pairs},t_{\pairs})\}$ of pairs of vertices, and an integer
  ${\size}$, the \textsc{Edge Multicut} problem ask if there is a set
  $S$ of at most ${\size}$ edges such that the removal of $S$
  disconnects every $s_i$ from the corresponding $t_i$. \textsc{Vertex
    Multicut} is the analogous problem where $S$ is a set of at most
  ${\size}$ vertices. Our main result is that both problems can be
  solved in time $2^{O({\size}^3)}\cdot n^{O(1)}$, i.e.,
  fixed-parameter tractable parameterized by the size ${\size}$ of the
  cutset in the solution. By contrast, it is unlikely
  that an algorithm with running time of the form $f({\size})\cdot
  n^{O(1)}$ exists for the directed version of the problem, as we show
  it to be
  W[1]-hard parameterized by the size of the cutset.
\end{abstract}

\section{Introduction}
From the classical results of Ford and Fulkerson on minimum $s-t$ cuts \cite{MR0079251}
to the more recent $O(\sqrt{\log n})$-approximation algorithms for
sparsest cut problems \cite{DBLP:conf/focs/Sherman09,1502794,DBLP:journals/siamcomp/FeigeHL08}, the
study of cut and separation problems have a deep and rich theory.
One well-studied problem in this area is the \textsc{Edge Multicut}
problem: given a graph $G$ and pairs of vertices $(s_1,t_1)$,
$\dots$, $(s_{\pairs},t_{\pairs})$, remove a minimum set of edges such that every
$s_i$ is disconnected from its corresponding $t_i$ for every $1\le
i\le {\pairs}$. For ${\pairs}=1$, \textsc{Edge Multicut} is the classical $s-t$ cut
problem and can be solved in polynomial time. For ${\pairs}=2$, \textsc{Edge Multicut} remains
polynomial-time solvable \cite{DBLP:conf/icalp/YannakakisKCP83}, but
it becomes NP-hard for every fixed ${\pairs}\ge 3$ \cite{MR95h:90039}.
\textsc{Edge Multicut} can be approximated within a factor of $O(\log
{\pairs})$ in polynomial time \cite{MR1379299} (even in the weighted case
where the goal is to minimize the total weight of the removed edges).
However, under the Unique Games Conjecture of Khot
\cite{MR2121525-full}, no constant factor approximation is possible
 \cite{MR2243123}. One can analogously
define the \textsc{Vertex Multicut} problem, where the task is to
remove a minimum set of vertices. An easy reduction shows that the
vertex version is more general than the edge version.

Using brute force, one can decide in time $n^{O({\size})}$ if  a
solution of size at most ${\size}$ exists. Our main result is a more efficient exact
algorithm for small values of ${\size}$ (the $O^*$ notation hides
factors that are polynomial in the input size):
\begin{theorem}\label{th:main}
Given an instance of
\textsc{Vertex Multicut} or \textsc{Edge Multicut} and an integer ${\size}$, one can find 
in time $O^*(2^{O({\size}^3)})$ a solution of size ${\size}$, if such a solution exists.
\end{theorem}
That is, we prove that \textsc{Vertex Multicut} and \textsc{Edge
  Multicut} are fixed-parameter tractable parameterized by the size
${\size}$ of the solution, resolving a very challenging open question
in the area of parameterized complexity.
 (Recall that a problem
is {\em fixed-parameter tractable} (FPT) with a particular parameter
${\size}$ if it can be solved in time $f({\size})\cdot n^{O(1)}$,
where $f$ is an arbitrary computable function depending only on ${\size}$; see
\cite{MR2001b:68042,grohe-flum-param,Niederbook} for more background).
The question was first asked explicitly perhaps in
\cite{marx-separation}; it has been restated more recently as an open
problem in e.g., \cite{MR2363707, ChenMwaycutAlgorithmica}.
  Our result shows in particular
that multicut is polynomial-time solvable if the size of the optimum
solution is $O(\sqrt[3]{\log n})$ (where $n$ is the input size).

One reason why multicut is a fundamental problem is that it is able to
express several other problems.  It has been observed that a
correlation clustering problem called \textsc{Fuzzy Cluster Editing}
can be reduced to (and in fact, equivalent with) \textsc{Edge Multicut}
\cite{DBLP:journals/tcs/BodlaenderFHMPR10,DBLP:journals/tcs/DemaineEFI06,DBLP:journals/ml/BansalBC04}.
Our results show that \textsc{Fuzzy Cluster Editing}
is FPT parameterized by the editing cost, settling this open problem
discussed e.g., in \cite{DBLP:journals/tcs/BodlaenderFHMPR10}.

\textbf{Previous work.}  The fixed-parameter tractability of
multicut and related problems has been thoroughly investigated in the
literature. \textsc{Edge Multicut} is NP-hard on trees, but it is
known to be FPT, parameterized by the maximum number ${\size}$ of edges that
can be deleted, and admits a polynomial kernel
\cite{stacs-poly-kernel,MR2165486}. Multicut problems were studied in
\cite{MR2363707} for certain restricted classes of graphs.  For
general graphs, \textsc{Vertex Multicut} is FPT if both ${\size}$ and and
the number of terminal pairs ${\pairs}$ are chosen as parameters (i.e, the
problem can be solved in time $f({\size},{\pairs})\cdot n^{O(1)}$
\cite{MarxTCS,DBLP:conf/csr/Xiao08,Guillemot2010} for some function $f$). The algorithm of
Theorem~\ref{th:main} is superior to these result in the sense that
the running time depends polynomially on the number ${\pairs}$ of terminals pairs, and
the exponential dependence is restricted to the parameter ${\size}$, the number of deletions. For the
special case of \textsc{Multiway Cut} (where terminals in a set $T$
have to be pairwise separated form each other), algorithms with
running time of the form $f({\size})\cdot n^{O(1)}$ were already known
\cite{MarxTCS,ChenMwaycutAlgorithmica,Guillemot2010}, but apparently these
algorithms do not generalize in an easy way to multicut.  An
FPT 2-approximation
algorithm was given in \cite{DBLP:journals/ipl/MarxR09} for
\textsc{Edge Multicut}: in time $O^*(2^{O({\size}\log{\size})})$, one can find
a solution of size $2{\size}$ if a solution of size ${\size}$ exists. There is no
obvious FPT algorithm for the problem even on bounded-treewidth
graphs, although one can obtain linear-time algorithms if the
treewidth remains bounded after adding an edge $s_it_i$ for each
terminal pair
\cite{DBLP:journals/ipl/GottlobL07,springerlink:10.1007/978-3-642-13073-1}.
A  PTAS is known for bounded-degree graphs of bounded treewidth \cite{Calinescu2003333}.

\textbf{Our techniques.} The first two steps of our algorithm follows
\cite{DBLP:journals/ipl/MarxR09}. We start by an opening step that is
fairly standard in the design of FPT algorithms. Instead of solving the
original \textsc{Vertex Multicut} problem, we solve the compression
version of the problem, where the input contains a solution $W$ of
size ${\size}+1$, and the task is to find a solution of size ${\size}$
(if exists). A standard argument called {\em iterative compression}
\cite{Reed1,HuffnerSurvey} shows that if the compression problem is
FPT, then the original problem is FPT.  Alternatively, we can use the
polynomial-time approximation algorithm of Gupta~\cite{MR1974949},
which produces a solution $W$ of size ${\size}^2$ if a solution of
size ${\size}$ exists. In this case, $O({\size}^2)$ iterations of the
compression algorithm gives a solution of size ${\size}$.

Next, as in \cite{DBLP:journals/ipl/MarxR09}, we try to reduce the
compression problem to \textsc{Almost 2SAT} (delete $k$ clauses to
make a 2-CNF formula satisfiable; also known as \textsc{2CNF
  Deletion}), which is known to be FPT
\cite{DBLP:journals/jcss/RazgonO09,DBLP:journals/toct/CyganPPW13,DBLP:conf/esa/RamanRS11}.
However, our 2SAT formulation is very different from the one in
\cite{DBLP:journals/ipl/MarxR09}: we introduce a single variable $x_v$
only for each vertex of $G$, while in \cite{DBLP:journals/ipl/MarxR09}
there is a variable $x_{v,w}$ for every vertex $v\in V(G)$ and vertex
$w\in W$ of the initial solution. This simpler reduction to
\textsc{Almost 2SAT} is correct only if the instance satisfies two
quite special properties:
\begin{itemize}
\item[(1)] every component of $G\setminus W$ is adjacent to at most two vertices
of $W$ (``has at most two legs''), and 
\item[(2)] there is a
solution $S$ such that every component of $G\setminus S$ contains a
vertex of $W$ (``no vertex is isolated from $W$ after removing the 
solution $S$'' or ``no vertex is in the shadow of $S$'').
\end{itemize}
 The main part of the paper is devoted to showing how these
properties can be achieved.  In order to achieve
property (1), we show by an analysis of cuts and performing
appropriate branching steps that the set $W$ can be extended in such a way
that every component has at most two legs
(Section~\ref{sec:reduct-biped-case}). To achieve property (2), we
describe a nontrivial way of sampling random subset of vertices such
that if we remove this subset by a certain contraction operation
(taking the torso of the graph), then without changing the solution,
we get rid of the parts not reachable from $W$ with some positive
probability (Section~\ref{sec:creat-nonis-solut}).  This random
sampling uses the concept of ``important separators,'' which was
introduced in \cite{MarxTCS}, and has been implicitly used in
\cite{DBLP:journals/jacm/ChenLLOR08,DBLP:journals/jcss/RazgonO09,ChenMwaycutAlgorithmica}
in the design of parameterized algorithms. We consider the random
sampling of important separators the main new technical idea of the
paper. This technique and its generalizations have turned out to be useful for other problems
as well \cite{dirmwaysicomp,DBLP:journals/iandc/LokshtanovM13,DBLP:conf/icalp/ChitnisCHM12,dirmwaysicomp,esa13-homdel,DBLP:conf/icalp/LokshtanovR12,DBLP:conf/icalp/KratschPPW12} and we
expect it to have further application in the future.

\textbf{Directed graphs.} Having resolved the fixed-parameter
tractability of \textsc{Vertex Multicut}, the next obvious question is
what happens on directed graphs. Note that for directed graphs, the
edge and vertex versions are equivalent. In directed graphs, multicut
becomes much harder to approximate: there is no polynomial-time
$2^{\log ^{1-\epsilon}n}$-approximation for any $\epsilon>0$, unless
$\textup{NP}\subseteq \textup{ZPP}$ \cite{1502795}.  From the
fixed-parameter tractability point of view, the directed version of
the problem received particular attention because \textsc{Directed
  Feedback Vertex Set} or DFVS (delete ${\size}$ vertices to make the
graph acyclic) can be reduced to \textsc{Directed Multicut}. The
fixed-parameter tractability of DFVS had been a longstanding open
question in the area of parameterized complexity until it was solved
by Chen et al.~\cite{DBLP:journals/jacm/ChenLLOR08} recently.
The main idea that led to the solution is that DFVS can be reduced to
a variant (in fact, special case) of \textsc{Directed Multicut} called
\textsc{Skew Multicut}, where the task is to break every path from
$s_i$ to $t_j$ for every $i>j$. By showing that \textsc{Skew Multicut}
is FPT parameterized by the size of the solution, Chen et
al.~\cite{DBLP:journals/jacm/ChenLLOR08} proved the
fixed-parameter tractability of DFVS.
 We show in Section~\ref{sec:hardn-rm-textscd} that, unlike \textsc{Skew
  Multicut}, the general \textsc{Directed Multicut} problem is
unlikely to be FPT.
\begin{theorem}\label{th:dirhard}
\textsc{Directed Multicut} is \textup{W[1]}-hard parameterized by the
size ${\size}$ of the solution.
\end{theorem}

\textbf{Independent and followup work.} A preliminary version of this
paper appeared in \cite{marx-razgon-multicut}; the current version
contains essentially the same algorithm, but the terminology and
organization of Section~\ref{sec:find-nonis-solut} were significantly
changed.  Independently from our work, Bousquet et
al.~\cite{dblp:conf/stoc/bousquetdt11} presented in the same volume a
proof that \textsc{Multicut} is FPT parameterized by the size
${\size}$ of the solution. The two algorithms have certain parts in
common: both reduce the problem to the compression version and both
ensure that we have to deal with components having only two
legs. However, the main part of the two algorithms are substantially
different: the current paper introduces the technique of random
sampling of important separators and uses it to reduce the problem to
\textsc{Almost 2SAT}, while Bousquet et
al.~\cite{dblp:conf/stoc/bousquetdt11} uses an approach based on a
series of problem-specific reductions to reduce the problem to
\textsc{2SAT}.

Subsequently to the first version of this paper, random sampling of
important separators has been used in several other applications. For undirected graphs, the technique was used by Lokshtanov and Ramanujan~\cite{DBLP:conf/icalp/LokshtanovR12} to solve a parity version of \textsc{Multiway Cut} and by Chitnis et al.~\cite{esa13-homdel} to solve a homomorphism problem generalizing certain deletion problems. For directed graphs, even
though \textsc{Directed Multicut} is W[1]-hard parameterized by
${\size}$ (see Section~\ref{sec:hardn-rm-textscd}), Chitnis et
al.~\cite{dirmwaysicomp} proved that the special case \textsc{Directed
  Multiway Cut} (Given a set $T$ of terminals, break every directed
path between two different terminals by removing at most ${\size}$
edges/vertices) is FPT parameterized by ${\size}$. A consequence of
this result is that \textsc{Directed Multicut} with ${\pairs}=2$ is
FPT parameterized by ${\size}$ is FPT. Kratsch et al.~\cite{DBLP:conf/icalp/KratschPPW12} proved that \textsc{Directed Multicut} on directed acyclic graphs (DAGs) is FPT with combined parameters 
${\pairs}$ and ${\size}$, and strenghed our hardness result by showing that \textsc{Directed Multicut} remains W[1]-hard parameterized by ${\size}$ even on DAGs. However, the complexity of
\textsc{Directed Multicut} for ${\pairs}=3$ or with combined
parameters ${\pairs}$ and ${\size}$ remains an interesting open
question. 

Chitnis et al.~\cite{DBLP:conf/icalp/ChitnisCHM12} use the random
sampling technique to show the fixed-parmeter tractability of
\textsc{Directed Subset Feedback Vertex Set}. They present an abstract
framework that formalizes under which conditions this technique can
be used, and they improve the randomized selection and its analysis to
obtain better success probability and improved running time.

A very different application of the technique is given by Lokshtanov
and Marx~\cite{DBLP:journals/iandc/LokshtanovM13} in the context of
clustering problems. They study a family of clustering problems such as
partitioning the vertices of an undirected graph into clusters of size
at most $p$ such that at most $q$ edges leave each cluster. The
problem boils down to being able to check whether a given vertex $v$
is contained in such a cluster. It turns out that the random sampling
of important separators technique can be used to show that this task
(and therefore the original clustering problem) is FPT parameterized by $q$ by
reducing it to a knapsack-like problem.

\section{Framework: compression, shadows, legs}\label{sec:preliminaries}
\begin{figure}
\tikzstyle{block} = [outer sep=0.6em, rectangle, draw,
    text width=18em, text centered, rounded corners, minimum height=4em]
\tikzstyle{endblock} = [outer sep=1em, rectangle,
    text width=18em, text centered, rounded corners]
\tikzstyle{line} = [draw, -triangle 45]
\tikzstyle{arrownode} =[rectangle,text width=22em]
\begin{center}
{\centering
\small \begin{tikzpicture}[node distance = 4cm, auto]
    \node [block] (mc) {\mc};
    \node [block, below of=mc] (mcc) {\mccs};
    \node [block, below of=mcc] (smcc) {\mccs\\ (shadowless solution)};
    \node [block, below of=smcc] (bmcc) {\bmcc\\ (shadowless solution)};
    \node [block, below of=bmcc] (a2sat) {\textsc{Almost 2SAT}};
    \node [endblock, below of=a2sat] (fpt) {FPT};

    \path [line] (mc) -- node[arrownode] {Iterative compression or approximation\\(Section~\ref{sec:preliminaries} )}  (mcc);
    \path [line] (mcc) -- node[arrownode] {Random sampling of important separators\\(Section~\ref{sec:creat-nonis-solut})} (smcc);
    \path [line] (smcc) -- node[arrownode] {Branching on shattering sets\\(Section~\ref{sec:reduct-biped-case})} (bmcc);
    \path [line] (bmcc) -- node[arrownode] {Encoding into 2SAT\\(Section~\ref{sec:find-nonis-solut})} (a2sat);
    \path [line] (a2sat) -- node[arrownode]{Previous work\\\cite{DBLP:journals/jcss/RazgonO09,DBLP:journals/toct/CyganPPW13,DBLP:conf/esa/RamanRS11,DBLP:journals/corr/abs-1203-0833}} (fpt);
\end{tikzpicture}
}
\end{center}
\caption{The chain of reductions in the paper.}\label{fig:chain}
\end{figure}
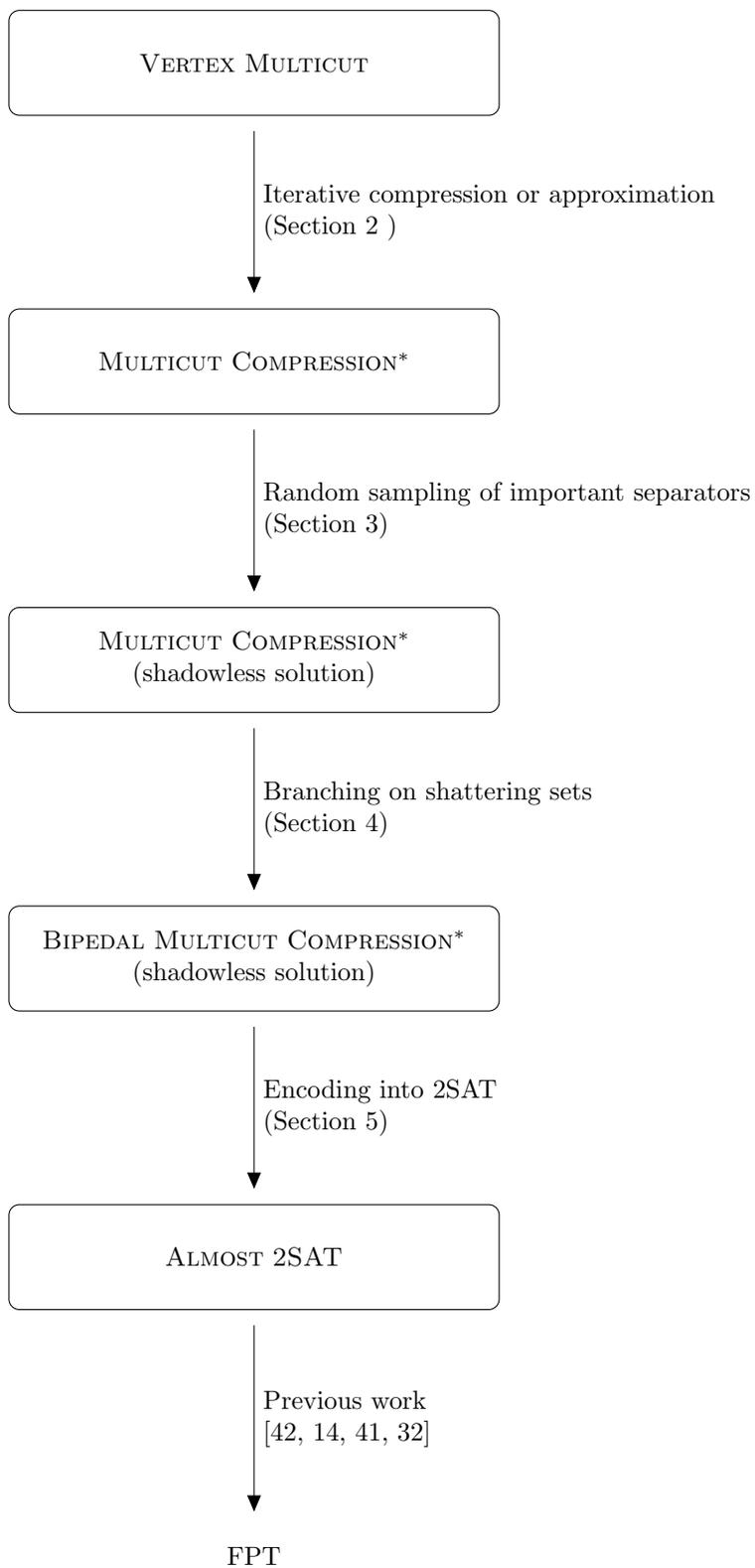

Let $G$ be an undirected graph and let ${\bf
  T}=\{(s_1,t_1),\dots,(s_{\pairs},t_{\pairs})\}$ be a set of terminal pairs. We say
that a set $S\subseteq V(G)$ of vertices is a {\em multicut} of
$(G,{\bf T})$ if there is no component\footnote{Throughout this paper, 
when we refer to a component $K$ of a graph, we consider
the set of vertices of this component. We omit saying ``the set of vertices of'' for the
sake of brevity.} of $G\setminus S$ that contains
both $s_i$ and $t_i$ for some $1\le i \le {\pairs}$ (note that it is
allowed that $S$ contains $s_i$ or $t_i$). The central problem of
the paper is the following:

\begin{center}
\fbox{
\parbox{0.9\linewidth}{
  \noindent{\mc}
\smallskip

\begin{tabular}{rl}
  \emph{Input:} &A graph $G$, an integer ${\size}$, and\\
& a set ${\bf T}$ of pairs of 
  vertices of $G$.\\
  \emph{Output:} &A multicut of $(G,{\bf T})$ of size at most ${\size}$\\&
 or ``NO'' if no
  such multicut exists.
\end{tabular}
}}
\end{center}
We prove the fixed-parameter tractability of \mc\ by a series of
reductions (see Figure~\ref{fig:chain}).  First we argue that it is
sufficient to solve an easier {\em solution compression} problem. Then
we present two reductions that modify the problem in such a way that it is
sufficient to look for solutions that are {\em shadowless} and
we can assume that the instance is {\em bipedal}. The last step of the proof is reducing this
special variant of the problem to \textsc{Almost 2SAT}.

\subsection{Compression}
The first step in the proof of Theorem~\ref{th:main} is a standard
technique in the design of parameterized algorithms: we define and
solve the {\em compression problem}, where it is assumed that the
input contains a feasible solution of size larger than ${\size}$. As
this technique is standard (and in particular, we follow the approach
of \cite{DBLP:journals/ipl/MarxR09} for \textsc{Edge Multicut}), we
keep this section short and informal.
\begin{center}
\fbox{
\parbox{0.9\linewidth}{
  \noindent{\mcc}
\smallskip

\begin{tabular}{rl}
  \emph{Input:} & A graph $G$, an integer ${\size}$,\\
& a set ${\bf T}$ of pairs of 
  vertices of $G$, and\\
& a multicut $W$ of $(G,{\bf T})$.\\
  \emph{Output:} & A multicut of $(G,{\bf T})$ of size at most ${\size}$,\\
& or ``NO'' if no
  such multicut exists.
\end{tabular}
}}
\end{center}
Our main technical contribution is showing that \mcc\ is
FPT parameterized by ${\size}$ and $|W|$.
\begin{lemma}\label{lem:compression}
\mcc\ can be solved in time $O^*(2^{O(({\size}+\log|W|)^3+|W|\log |W|)})$.
\end{lemma}
Intuitively, it is clear that proving Lemma~\ref{lem:compression}
could be easier than proving that \mc\ is FPT: the extra input $W$ can
give us useful structural information about the graph (and as $|W|$
appears in the running time, a large $W$ is also helpful).
What's not obvious is how solving \mcc\ gives us any
help in the solution of the original \mc\ problem. We sketch two methods.

\textbf{Method 1.} Let us use the polynomial-time approximation
algorithm of Gupta~\cite{MR1974949} to find a multicut $W$ of size at most
$c\cdot \textup{OPT}^2$, where $c$ is a universal constant and $\textup{OPT}$ is the
minimum size of a multicut. If $|W|\ge c\cdot {\size}^2$, then we can safely
answer ``NO'', as there is no multicut of size at most ${\size}$. Otherwise,
we run the algorithm of Lemma~\ref{lem:compression} for this set $W$
to obtain a solution in time
$O^*(2^{O(({\size}+\log|W|)^3)}=O^*(2^{O({\size}^3)})$.

\textbf{Method 2.} The standard technique of {\em iterative
  compression} \cite{Reed1,HuffnerSurvey} allows us to reduce \mc\ to
at most $|V(G)|$ instances of \mcc\ with $|W|={\size}+1$. This
technique was used for the 2-approximation of \textsc{Edge Multicut}
in \cite{DBLP:journals/ipl/MarxR09} and its application is analogous
in our case.  Let $(G,{\bf T},{\size})$ be an instance of \mc. Suppose
that $V(G)=\{v_1, \dots, v_n\}$, let $G_i=G[\{v_1,\dots,v_i\}]$, and
let ${\bf T}_i$ be the subset of ${\bf T}$ containing the pairs with
both endpoints in $G_i$.  One by one, we consider the instances
$(G_i,{\bf T_i},{\size})$ in ascending order of $i$, and for each
instance we find a solution $S_i$ of size at most ${\size}$. We start
with $S_0=\emptyset$. For some $i>0$, we compute $S_i$ provided that
$S_{i-1}$ is already known. Observe that $S_{i-1}\cup \{v_i\}$ is a
multicut of size at most ${\size}+1$ for $(G_i,{\bf T}_i)$. Thus we can use the
algorithm for \mcc, which either returns a multicut $S_i$ of
$(G_i,{\bf T}_i)$ having size at most ${\size}$ or returns ``NO''. In
the first case, we can continue the iteration with $i+1$. In the
second case, we know that
there is no multicut of size ${\size}$ for $(G,{\bf T})$ (as there is
no such multicut even for $(G_i,{\bf T}_i)$), and hence we can return
``NO''.

Both methods result in $O^*(2^{O({\size}^3)})$ time 
algorithms. However, we feel it important to mention both approaches, as
improvements in Lemma~\ref{lem:compression} might have different
effects on the two methods.

It will be convenient to work with a slightly modified version of the
compression problem. We say that a set $S\subseteq V(G)$ is a {\em
  multiway cut} of $W\subseteq V(G)$ if every component of $G\setminus
S$ contains at most one vertex of $W$.
\begin{center}
\fbox{
\parbox{0.9\linewidth}{
  \noindent{\mccs}
\smallskip

\begin{tabularx}{\linewidth}{rX}
  \emph{Input:} &A graph $G$, an integer ${\size}$, \\
&a set ${\bf T}$ of pairs of 
  vertices of $G$, and
\\& a multicut $W$ of $(G,{\bf T})$.\\
  \emph{Output:} &A set $S$ of size at most ${\size}$ such that
\begin{enumerate}
\item[(1)]  $S$ is
  multicut of $(G,{\bf T})$,
\item[(2)] $S\cap
  W=\emptyset$, and
\item[(3)] $S$ is a multiway cut of $W$
\end{enumerate}
or ``NO'' if no
  such set $S$ exists.
\end{tabularx}
}}
\end{center}
That is, \mccs\ has two additional constraints on the solution $S$. 
In Sections~\ref{sec:reduct-biped-case}--\ref{sec:find-nonis-solut},
we prove that this problem is FPT:
\begin{lemma}\label{lem:compression2}
\mccs\ can be solved in time $O^*(2^{O(({\size}+\log|W|)^3)})$.
\end{lemma}
It is not difficult to reduce \mcc\ to \mccs\ (an analogous reduction
 was done in \cite{DBLP:journals/ipl/MarxR09} for the
the edge case). We briefly sketch such a reduction. In order to
solve an instance $(G,{\bf T},W,{\size})$ of \mcc, we first guess the intersection
$X$ of the multicut $W$ given in the input and the solution $S$ we are
looking for. This guess results in at most 
$\sum_{i=1}^{{\size}}\binom{|W|}{i}$ branches; in each
branch, we remove the vertices of $X$ from $G$ and decrease ${\size}$ by
$|X|$. Thus in the following, we can restrict our attention to
solutions disjoint from $W$. Next, we branch on all possible
partitions $(W_1,\dots,W_t)$ of $W$, contract each $W_i$ into a single
vertex, and solve \mccs\ on the resulting instance $(G',{\bf T}',W',{\size}')$. One of the
partitions $(W_1,\dots,W_t)$ corresponds to the way the solution $S$
partitions $W$ into connected components, and in this case $S$ is a
multiway cut of $W'$ in $G'$. Thus if the original \mcc\ instance has
a solution $S$, then it is a solution of one of the constructed \mccs\
instances. Conversely, any solution of the constructed instances is a
solution of the original instance. As the number of partitions of
$W$ can be bounded by $|W|^{O(|W|)}$, the running time claimed in
Lemma~\ref{lem:compression} follows from Lemma~\ref{lem:compression2}.
Thus in the rest of the paper, it is sufficient to prove
Lemma~\ref{lem:compression2} to obtain the main result, i.e.,
Theorem~\ref{th:main}.
Thus proving Lemma~\ref{lem:compression2} implies the main result Theorem~\ref{th:main}.

\subsection{Shadows}
\begin{figure}
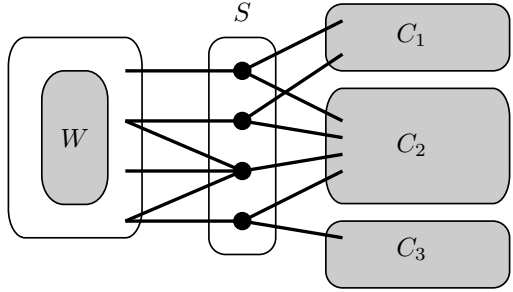

\begin{center}
{\small \svg{0.4\linewidth}{shadow}}
\caption{The shadow of $S$ consists of the three sets $C_1$, $C_2$, $C_3$.}\label{fig:shadow}
\end{center}
\end{figure}

An important step in our algorithm for \textsc{Multicut} (and in further applications of the randomized sampling of important separators method) is to argue about solutions that are  ``shadowless'' in the sense defined below. Intuitively, we imagine the vertices in $W$ as light
sources, light spreads on the edges, and $S$ blocks the light (see Figure~\ref{fig:shadow}).
\begin{defin}
Let $I=(G,{\bf T},W,{\size})$ be an instance of the \mccs\ problem, and
let $S$ be a solution for $I$.  The {\em shadow} of the set $S$ is
the set of vertices not reachable from any vertex of $W$ in
$G\setminus S$.  We
say that the solution $S$ is {\em shadowless} if the shadow is empty,
i.e., $G\setminus S$ has exactly $|W|$ components. 
\end{defin}
In Section~\ref{sec:creat-nonis-solut}, we present a randomized
algorithm that modifies the instance such that if a solution exists,
then it makes the solution shadowless with positive probability. The
algorithm is based on a randomized contraction of sets defined by
``important separators''; we review this concept in
Section~\ref{sec:important-separators}.  The algorithm can be
derandomized to obtain the following lemma:
\begin{lemma}[shadowless reduction]\label{lem:shadowless}  
Given an instance $I$ of the \mccs\ problem, we can construct in  time $O^*(2^{O({\size}^3)})$ a set of $t=2^{O({\size}^3)}\log n$ instances $I_1$, $\dots$, $I_t$, 
each with the same parameter ${\size}$ as $I$, such that
\begin{enumerate}
\item Any solution of $I_i$ for any $1\le i \le t$ is a solution of $I$.
\item If $I$ has a solution, then $I_i$ has a {\em shadowless} solution
  for at least one $1\le i \le t$.
\end{enumerate}
\end{lemma}
Thus Lemma~\ref{lem:shadowless} allows us to reduce the \mccs\ problem
into a variant where the task is to find a shadowless solution.

\subsection{Components and legs}
\begin{figure}
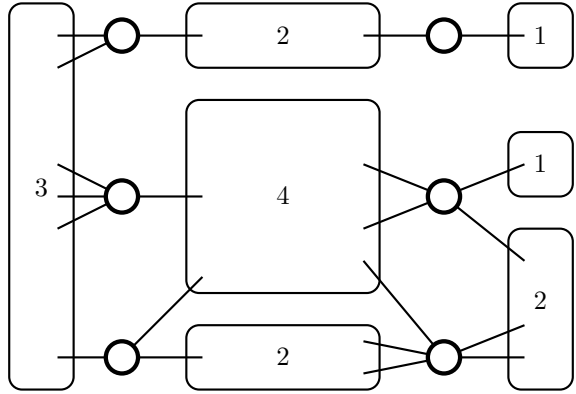

\begin{center}
{\small \svg{0.45\linewidth}{legs}}
\caption{An instance with 7 components. The strong circles are the
  vertices of $W$, the numbers show the number of legs for each component.}\label{fig:legs}
\end{center}
\end{figure}

In order to find a shadowless solution for a \mccs\ instance, the
problem is further transformed in Section~\ref{sec:reduct-biped-case}
using the concept of \emph{legs}.
\begin{defin}  Given an instance $(G,{\bf
  T},W,{\size})$ of \mccs, we say that a component $C$ of $G\setminus
W$ has {\em $\ell$-legs} if $C$ is adjacent with $\ell$ vertices of
$W$ (see Figure~\ref{fig:legs}). We say that a \mccs\ instance is {\em bipedal} if every component
of $G\setminus W$ has at most two legs; \bmcc\ is the problem
restricted to such instances.
\end{defin}
The transformation presented in
Section~\ref{sec:reduct-biped-case} reduces \mccs\ to a bounded number
of bipedal instances.
\begin{lemma}[bipedal reduction]\label{lem:shatterit} 
Given an instance $I$ of the \mccs\ problem with parameter ${\size}$, in  time $O^*(2^{O(({\size}+\log|W|)^3)})$ we can 
either solve this instance or construct a set of $t=2^{O({\size}+\log|W|)^3}$  instances $I_1$, $\dots$, $I_t$, 
of \bmcc\, each with parameter at most ${\size}$, such that
\begin{enumerate}
\item Any solution of $I_i$ for any $1\le i \le t$ is a solution of $I$.
\item If $I$ has a shadowless solution, then $I_i$ has a shadowless
  solution for at least one $1\le i \le t$.
\end{enumerate}
\end{lemma}
Finally, in Section~\ref{sec:find-nonis-solut}, we show 
how this solution can be found by a quite 
intuitive reduction to an FPT problem \textsc{Almost 2SAT}. 

\begin{lemma}\label{lem:shadowless2sat}
  Let $I=(G,{\bf T},W,{\size})$ be an instance of \bmcc\ that has a shadowless solution $S$
of size at most $\size$. In
time $O^*(4^{\size})$, we can find a (not necessarily shadowless)
solution $S'$.
\end{lemma}

Combining Lemmas~\ref{lem:shadowless}--\ref{lem:shadowless2sat} allows
us to prove Lemma \ref{lem:compression2} and therefore to solve
\mc.

\begin{proof}[Proof (of Lemma \ref{lem:compression2}).]
  Let us apply the Algorithm of Lemma \ref{lem:shadowless} to an
  instance $I=(G,{\bf T},W,{\size})$ of \mccs. This algorithm takes
  time $O^*(2^{O({\size}^3)})$ and produces $t=2^{O({\size}^3)}\log n$
  instances $I_i$ of the \mccs\ problem, each with parameter at most
  ${\size}$, so that the original instance $I$ has a solution if and
  only if one of these $t$ instances has a \emph{shadowless}
  solution. Moreover a (not necessarily shadowless) solution of any of
  these instances is also a solution of the orginal instance.

  Apply to each instance $I_i$ the algorithm of Lemma
  \ref{lem:shatterit}, which in time $O^*(2^{O(({\size}+\log|W|)^3)})$
  either returns an answer or produces $2^{O(({\size}+\log|W|)^3)}$
  instances $I_{i,j}$, each with parameter at most ${\size}$, of the
  \bmcc\ problem such that $I_i$ has a shadowless solution if and only
  if at least one $I_{i,j}$ has a shadowless solution. Moreover a (not
  necessarily shadowless) solution of any new instance $I_{i,j}$ is also a solution of $I_i$.

Combining the above two steps, we conclude that in time $O^*(2^{O(({\size}+\log|W|)^3)})$ the algorithm produces 
$2^{O(({\size}+\log|W|)^3)}\log n$ instances  of the \bmcc\ problem such that
the original instance $I$   has a solution if and only if at least one of the
these $2^{O(({\size}+\log|W|)^3)}\log n$ instances has a shadowless solution. Moreover a (not necessarily shadowless)
solution of any instance $I_{i,j}$ is also a solution of $I$.

Finally, we apply to each resulting instance $I_{i,j}$ of the \bmcc\ problem the
algorithm of Lemma~\ref{lem:shadowless2sat}. By the discussion above, if the algorithm returns
a solution for at least one of the instances, then this
is a solution of the original instance $I$. If the algorithm returns
``NO'' for all the instances, this means that no one of them has a
shadowless solution. It follows that the original instance does not
have a solution either. Taking into account that the algorithm of
Lemma~\ref{lem:shadowless2sat} takes time $O^*(4^{\size})$, processing
of $2^{O(({\size}+\log|W|)^3)}\log n$ instances takes time
$O^*(2^{O(({\size}+\log|W|)^3)})$.  Consequently, the instance $I$ of
the \mccs\ problem can be solved in time
$O^*(2^{O(({\size}+\log|W|)^3)})$.
\end{proof}

\section{Making the solution shadowless}
\label{sec:creat-nonis-solut}

The purpose of this section is to reduce solving \mccs\ to finding a shadowless solution. 
We present a randomized transformation that, given an
instance having a solution, modifies the instance
in such a way that the new instance has a {\em shadowless}
solution with probability $2^{-O({\size}^3)}$.
More precisely:
\begin{lemma}\label{lem:shadowlessrand}
Given an instance $I$ of the \mccs\ problem, we can construct in  time $O^*(2^{O({\size})})$ an instance $I'$ 
with the same parameter ${\size}$ as $I$ such that
\begin{enumerate}
\item Any solution of $I'$ is a solution of $I$.
\item If $I$ has a solution, then $I'$ has a {\em shadowless} solution with probability $2^{-O({\size}^3)}$.
\end{enumerate}
\end{lemma}
This means that if $I$ has a solution, then by invoking
Lemma~\ref{lem:shadowlessrand} $2^{O({\size}^3)}$ times, with constant
probability at least one of the instances has a shadowless
solution. Thus if we are able to solve the problem with the assumption
that a shadowless solution exists, then this way we can get a solution
for $I$ with constant probability.
The main result of this
section is a derandomized version of this transformation (Lemma~\ref{lem:shadowless}).

The main idea in the proof of Lemma~\ref{lem:shadowlessrand} is to try to
randomly guess a set $Z$ whose removal does not change the instance
substantially, but makes the instance
shadowless. Section~\ref{sec:tors-shad-solut} introduces the torso
operation, which is used to remove the set $Z$, and states what
properties the set $Z$ needs to satisfy. The construction of $Z$ is
based on the observation that the solution can be characterized by a
``closest set'' and we need to locate the boundary of such a set
(Section~\ref{sec:closest-sets}). We develop a randomized algorithm
for this purpose in
Sections~\ref{sec:important-separators}--\ref{sec:derandomization}.
The algorithm uses the notion of important separators;
Section~\ref{sec:important-separators} reviews this concept and shows
why it is relevant for our
problem. Sections~\ref{sec:rand-select-sets}--\ref{sec:rand-select-sets-full}
describe and analyze the randomized selection
process. Section~\ref{sec:derandomization} shows how the random
selection can be derandomized to obtain the deterministic version, Lemma~\ref{lem:shadowless}.

\subsection{Torsos and shadowless solutions}\label{sec:tors-shad-solut}
The randomized transformation can be conveniently described using the
operation of taking the {\em torso} of a graph.
\begin{definition} \label{torsodef}
Let $G$ be a graph and $C\subseteq V(G)$. The graph $\torso(G,C)$ has
vertex set $C$ and two vertices $a,b\in C$ are adjacent if
$\{a,b\} \in E(G)$ or there is a path $P$ in $G$ connecting $a$ and $b$ whose internal
vertices are not in $C$.
\end{definition}
\begin{figure}
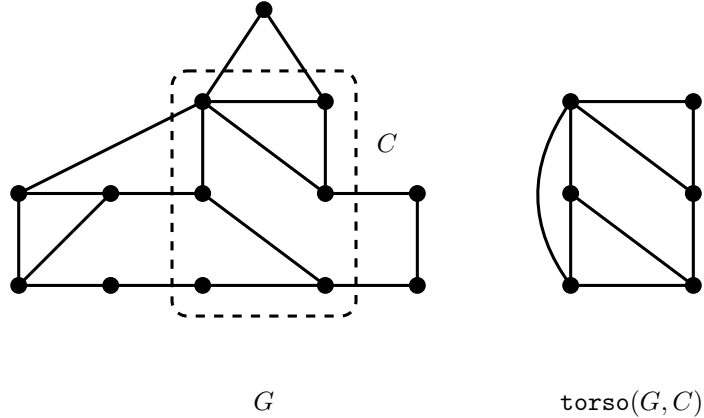

\begin{center}
{\small \svg{0.55\linewidth}{torso}}
\caption{The torso operation on the graph $G$ with a set $C$ of 6 vertices.}\label{fig:torso}
\end{center}
\end{figure}
In particular, every edge of $G[C]$ is in $\torso(G,C)$.  It is easy
to show that this operation preserves separation inside $C$:
\begin{proposition}\label{prop:torsosep}
  Let $C\subseteq V(G)$ be a set of vertices in $G$ and  let
  $a,b\in C$ two vertices. A set $S\subseteq C$ separates vertices $a$ and
  $b$ in $\torso(G,C)$ if and only if $S$ separates these vertices
  in $G$.
\end{proposition}
\begin{proof}
Let $P$ be a path connecting $a$ and $b$ in $G$ and
suppose that $P$ is disjoint from the set $S$. The path $P$ contains
vertices from $C$ and from $V(G)\setminus C$. If $u,v\in C$ are
two vertices such that every vertex of $P$ between $u$ and $v$ is from
$V(G)\setminus C$, then by definition there is an edge $uv$ in
$\torso(G,C)$. Using these edges, we can modify $P$ to obtain a path
$P'$ that connects $a$ and $b$ in $\torso(G,C)$ and avoids $S$.

Conversely, suppose that $P$ is a path connecting $a$ and $b$ in
the graph $\torso(G,C)$ and it avoids $S\subseteq C$. If $P$ uses an edge $uv$
that is not present in $G$, then this means that there is a path
connecting $u$ and $v$ whose internal vertices are not in $C$. Using
these paths, we can modify $P$ to obtain a path $P'$ that uses only
the edges of $G$. Since $S\subseteq C$, the new vertices on the path
are not in $S$, i.e., $P'$ avoids $S$ as well.
\end{proof} 

Let $I=(G,W,{\bf T},{\size})$ be an arbitrary instance of \mccs. Given a set
$Z\subseteq V(G)\setminus W$ of vertices, the  {\em reduced instance}
$\redi{Z}=(G',W,{\bf T}',{\size})$ is defined the following way:
\begin{enumerate}
\item The graph $G'$ is $\torso(G,V(G)\setminus Z)$.
\item For every $v\in V(G)$, let $\phi(v)=N(C)$ if $v$ belongs to
  component $C$ of $G[Z]$, and let $\phi(v)=\{v\}$ if $v\not\in
  Z$. The set ${\bf T}'$ is obtained by by replacing every
pair  $(x,y)\in {\bf T}$ with the set of pairs $\{(x',y')\mid x'\in \phi(x),y'\in
  \phi(y)\}$.
\end{enumerate}

The main observation is that if we perform this torso operation for a
$Z$ that is sufficiently large to cover the shadow of a
hypothetical solution $S$ and sufficiently small to be disjoint
from $S$, then $S$ becomes a shadowless solution of $I/Z$.
Furthermore, the torso operation is ``safe'' in the sense that it does
 not make the problem easier, i.e, does  not create new solutions.
\begin{lemma}\label{lem:shadowlessremove}
Let $I=(G,{\bf T},W,{\size})$ be an instance of \mccs\ and let $Z\subseteq
V(G)\setminus W$ be a set of vertices.
\begin{enumerate}
\item[(1)] Every solution of $\redi{Z}$ is a solution of $I$.

\item[(2)] If $I$ has a solution $S$ such that $Z$ covers the shadow
  and $Z\cap S=\emptyset$, then $S$ is a shadowless solution of
  $\redi{Z}$.
\end{enumerate}
\end{lemma}
\begin{proof}
  Let $G$ and $G'=\torso(G,V(G)\setminus Z)$ be the graphs in instances $I$ and $\redi{Z}$,
  respectively. To prove the first statement, we show that if
  $S'\subseteq V(G')$ is a solution of $\redi{Z}$, then $S'$ is a
  solution of $I$ as well. Suppose that some pair $(x,y)$ of $I$ is
  not separated by $S'$. Let $P$ be a path in $G\setminus S'$ going
  from $x$ to $y$.  Let $x'$ and $y'$ be the first and last vertex of
  $P$ not in $Z$, respectively, and let $P'$ be the subpath of $P$
  from $x'$ to $y'$. (Note that $P$ cannot be fully contained in $Z$,
  as it contains at least one vertex of the multicut $W$.)  By the way $\redi{Z}$
  is defined, $(x',y')$ is a pair in $\redi{Z}$, hence $S'$
  separates $x'$ and $y'$ in $G'=\torso(G,C)$. Using
  Prop.~\ref{prop:torsosep} with $C=V(G)\setminus Z$, we get that $S'$
  separates $x'$ and $y'$ in $G$, which is in contradiction with the
  existence of the path $P$.  A similar argument shows that there is
  no path in $G\setminus S'$ that connects two vertices of $W$.

  For the second statement, suppose that $S$ is a solution of $I$ with
  $S\cap Z=\emptyset$. Let us show that $S$ is a solution of
  $\redi{Z}$ as well. Suppose that $S$ does not separate $x'$ and $y'$
  in $G'$ for some pair $(x',y')$ of $\redi{Z}$. Using Prop.~\ref{prop:torsosep} with $C=V(G)\setminus
  Z$, we get that $S'$ does not separate $x'$ and $y'$ in $G$, i.e., there is
  an $x'-y'$ path $P$ in $G\setminus S$. By the way the
  pairs in $\redi{Z}$ were defined, there is a pair
  $(x,y)$ of $I$ and there is an $x-x'$ path $P_1$
  such that $x'$ is the only vertex of $P_1$ not in $Z$, and there is
  a $y-y'$ path $P_2$  such that $y'$ is the only vertex of $P_2$ not
  in $Z$. Clearly, these paths are disjoint form $S$. Therefore, the
  concatenation of $P_1$, $P$, $P_2$ is an $x-y$ path in $G\setminus
  S$, contradicting that $S$ is a solution of $I$.

  To see that $S$ is shadowless in $G'$, consider a vertex $v$ of
  $G'\setminus S$.  As $v\not\in Z$ is not in the shadow of the
  solution $S$ of $I$, there is a path $P$ in $G\setminus S$ going
  from $v$ to a vertex $w\in W$. Again by Prop.~\ref{prop:torsosep}, this
  means that there is a $v-w$ path in $G'\setminus S$ as well, which
  means that $v$ is not in the shadow of the solution $S$ of
  $I'$.
 \end{proof}

\subsection{Closest sets}
\label{sec:closest-sets}
Lemma~\ref{lem:shadowlessremove} shows that in order to reduce the
\mccs\ problem to finding a shadowless solution, all we need is a set
$Z$ that covers the shadow of a hypothetical solution $S$, but
disjoint from $S$ itself. It is not obvious how this observation is of
any help: it seems that there is no way of constructing such a set
without actually knowing  a solution $S$. Nevertheless, we present a
randomized procedure that constructs such a set with non-negligible
probability. 

The main idea of the randomized procedure is that a solution of a
\mccs\ instance can be characterized by the set of vertices reachable
from $W$, and we can assume that this set has the property that it
cannot be made smaller without increasing the size of the
boundary. The following definition formalizes this property:

\begin{definition}
Let $G$ be an undirected graph and let $W\subseteq V(G)$ be a subset of vertices. We say that a set $R\supseteq W$ is a {\em $W$-closest set} if there is no $R'\subset R$ with $R'\supseteq W$ and $|N(R')|\le |N(R)|$.
\end{definition}

The main technical idea of the paper is the following randomized
procedure, which, in some sense, finds the boundary of a closest
set. Note that this statement could be of independent interest, as it
is about closest sets in general and contains nothing specific to
multicut problems.
\begin{theorem}[random sampling]\label{th:randreduceabstract} There
  is a randomized algorithm $\randset(G,W,{\size})$ that, given a
  graph $G$, a set $W\subseteq V(G)$, and an integer ${\size}$,
  produces a set $Z\subseteq V(G)\setminus W$ such that the following
  holds. For every $W$-closest set $R$ with $|N(R)|\le {\size}$, the
  probability that the following two events both occur is at least
  $2^{-O({\size}^3)}$:
\begin{enumerate}
\item $N(R)\cap Z=\emptyset$, and
\item $V(G)\setminus (R\cup N(R))\subseteq Z$.
\end{enumerate}
\end{theorem}
That is, the two events say that $Z$ covers every vertex outside
$R\cup N(R)$ and may cover some vertices inside $R$, but disjoint from
$N(R)$.  To prove Theorem~\ref{th:randreduceabstract}, we introduce
the main new technique of the paper: random sampling of important
separators. In Section~\ref{sec:important-separators}, we review the
notion of important separators.  Section~\ref{sec:rand-select-sets}
contains a simplified proof of Theorem~\ref{th:randreduceabstract}
(with probability bound $2^{-2^{O({\size})}}$ instead of
$2^{-O({\size}^3)}$). The full proof appears in
Section~\ref{sec:rand-select-sets-full}.  We show below that
Theorem~\ref{th:randreduceabstract} can be used to prove
Lemma~\ref{lem:shadowlessrand}. Section~\ref{sec:derandomization}
shows how to derandomize Theorem~\ref{th:randreduceabstract}, which
immediately proves Lemma~\ref{lem:shadowless}.

\begin{proof}[Proof (of Lemma~\ref{lem:shadowlessrand}).]
  Let $I=(G,W,{\bf T},{\size})$ be an instance of \mccs.  Let us use
  the algorithm $\randset(G,W,{\size})$ of
  Theorem~\ref{th:randreduceabstract} to obtain a set $Z$ and let
  $I'=I/Z$. By Lemma~\ref{lem:shadowlessremove}, every solution of $I'$ is a solution of $I$ as well.

  Assume now that $I$ has a solution $S$; let $S$ be a solution such
  that $|S|$ is minimum possible, and among such solutions the set $R$
  of vertices reachable from $W$ in $G\setminus S$ is as small as
  possible. Clearly, $N(R)\subseteq S$. We claim that $R$ is a
  $W$-closest set. Suppose that there is a set $R'\subset R$
  containing $W$ such that $|N(R')|\le |N(R)|$. Let $S'=N(R')$, we
  have that $|S'|\le |S|$. We claim that $S'$ is a solution,
  contradicting the minimality of $S$. Suppose that there is a path
  $P$ in $G\setminus S'$ connecting the two terminals in a pair
  $(x,y)\in {\bf T}$ or two vertices of $W$. In both cases, $P$ has to
  go through a vertex of $W$ (here we use that the definition of
  $\mccs$ requires that $W$ is a multicut). Therefore, $P$ is fully
  contained in $R'\subset R$, which implies that it is disjoint from
  $N(R)\subseteq S$, i.e., $S$ is not a solution. Thus $S'$ is indeed a
  solution with $|S'|\le |S|$ and $|R'|<|R|$, contradicting the choice
  of the solution $S$. This contradiction proves our claim that $R$ is
  a $W$-closest set. The same argument shows that $N(R)$ is a
  solution, hence $S=N(R)$ has to hold.

  As $R$ is a $W$-closest set, the probability that both $S\cap
  Z=\emptyset$ and $V(G)\setminus (R\cup S)\subseteq Z$ hold is
  $2^{-O({\size}^3)}$. The later inclusion is equivalent to saying that
  the shadow of the solution $S$ is contained in $Z$. Therefore, by
  Lemma~\ref{lem:shadowlessremove}, set $S$ is a shadowless solution
  of instance $I'$.
\end{proof}

\subsection{Important separators}
\label{sec:important-separators}
The concept of important separators was introduced in \cite{MarxTCS}
to deal with the multiway cut problem. 
\begin{defin}
Let $G$ be an undirected graph and let $X,Y\subseteq V(G)$ be two
disjoint sets. A set
$S\subseteq V(G)$ of vertices is an {\em $X-Y$ separator} if $S$ is
disjoint from $X\cup Y$ and there is
no component $K$ of $G\setminus S$ with both $K\cap X\neq \emptyset$ and
$K\cap Y\neq \emptyset$. 
\end{defin}
In other words, $G\setminus S$ contains no path between $X$ and
$Y$. To improve readability, we write $s-Y$ separator instead of
$\{s\}-Y$ separator if $s$ is a single vertex. We emphasize the fact
that, by our definition, an $X-Y$ separators is disjoint from $X$ and
$Y$.

\begin{definition}\label{def:important}
Let $X,Y\subset V(G)$ be disjoint sets of vertices, $S\subseteq V(G)$ be an $X-Y$ separator, and
let $K$ be the union of every component of $G\setminus S$ intersecting
$X$. We say that
$S$ is an {\em important $X-Y$ separator} if it is inclusionwise
minimal and there is no $X-Y$
separator $S'$ with $|S'|\le |S|$ such that $K'\supset K$, where $K'$
is the union of every component of $G\setminus S'$ intersecting $X$.
\end{definition}
\begin{figure}
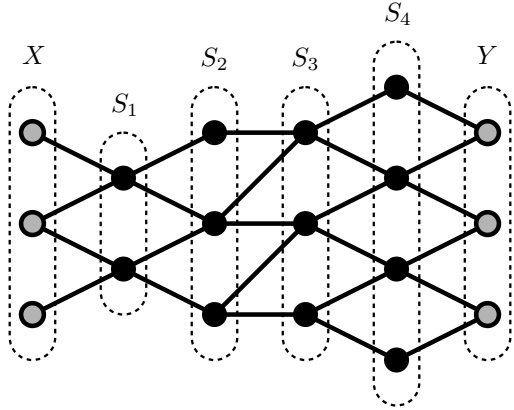

\begin{center}
{\small \svg{0.4\linewidth}{imp}}
\caption{Set $S_1$ is the unique minimum $X-Y$ separator and therefore it is an important $X-Y$ separator. Set $S_2$ is not an important $X-Y$ separator, as $|S_2|=|S_3|$ and a superset of vertices is reachable from $X$ in $G\setminus S_3$ compared to $G\setminus S_2$. Sets $S_3$ and $S_4$ are both important $X-Y$ separators.}\label{fig:imp}
\end{center}
\end{figure}
See Figure~\ref{fig:imp} for illustration.  Note that the order of
$X$ and $Y$ matters: an important $X-Y$ separator is not necessarily
an important $Y-X$ separator. It is easy to see that if $S$ is an
important $X-Y$ separator, then $S=N(R)$ for some set $R$ with
$X\subset R$ and $(R\cup N(R))\cap Y=\emptyset$: we can define 
$R$ to be the set of vertices reachable from $X$ in $G\setminus
S$. Observe that if $R$ is defined this way, then every component of
$G[R]$ contains at least one vertex of $X$. In particular, if $X$
contains only a single vertex, then we can assume that $G[R]$ is
connected.

A bound on the number of important separators was given in \cite{MarxTCS}
(although the notation there is slightly different). A better bound is
implicit in \cite{ChenMwaycutAlgorithmica}. 
For the convenience of the reader, we
give a self-contained proof of the following fact in the appendix.

\begin{lemma}\label{lem:impsep}
Let $X,Y\subseteq V(G)$ be disjoint sets of  vertices in a graph $G$. For every
${\size}\ge 0$, there are at most $4^{\size}$ important $X-Y$
separators of size at most ${\size}$. Furthermore, we can enumerate
all these separators in time $4^{\size}\cdot {\size} \cdot (|E(G)|+|V(G)|)$.
\end{lemma}

Note that one can give an exponential lower bound on the number of
important separators as a function of ${\size}$ and in fact
the bound $4^{\size}$ in Lemma~\ref{lem:impsep} is asymptotically tight up to factors polynomial in
${\size}$.

The following lemma connects closest sets and important separators by
showing that the boundary of a closest set is formed by important
separators. Intuitively, every vertex $v$ outside the closest set $R$
``sees'' a part of the boundary $N(R)$ that is an important $v-W$
separator: otherwise, we could ``push'' this part of the boundary away
from $v$ and towards $W$, contradicting the assumption that $R$ is a
closest set.
\begin{figure}
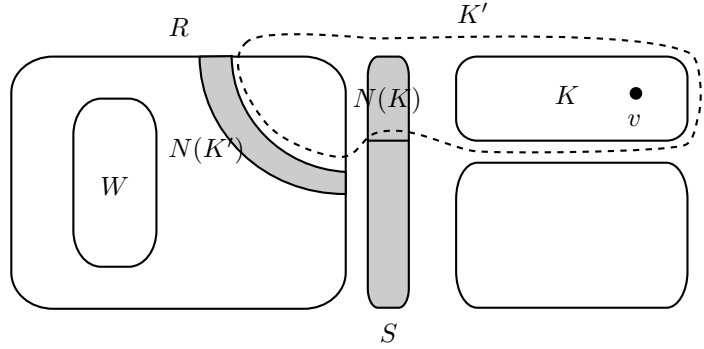

\begin{center}
{\small \svg{0.55\linewidth}{push}}
\caption{Proof of Lemma~\ref{lem:closestboundary}. Note that, in general, $K'$ can intersect other components of $G\setminus (R\cup N(R))$.}\label{fig:push}
\end{center}
\end{figure}
\begin{lemma}[pushing]\label{lem:closestboundary} Let $G$ be an
  undirected graph, $W$ a set of vertices, and $R$ a $W$-closest set.
  For every vertex $v\not\in R\cup N(R)$, there is an important $v-W$
  separator $S_v\subseteq N(R)$.
\end{lemma}
\begin{proof}
  Let $v$ be an arbitrary vertex of $G$ not in $R\cup N(R)$ and let
  $K$ be the component of $G\setminus N(R)$ containing $v$.  As
  $v\not\in R\cup N(R)$ and $W\subseteq R$, we have that $K$ is
  disjoint from $W$.  We show that $N(K)$ is an important $v-W$
  separator.  First, we observe that $N(K)$ is a minimal $v-W$
  separator: we have $N(K)\subseteq N(R)$, thus every vertex of $N(K)$
  is adjacent to both $K$ and $R$. Thus, if $N(K)$ is not an important
  $v-W$ separator, then there is a $K'\supset K$ such that $K'\cup N(K')$ is disjoint from $W$ and $|N(K')|\le
  |N(K)|$. We may assume that $G[K']$ is connected. Let
  $R':=R\setminus (K'\cup N(K'))$. Now $N(R')\subseteq (N(R)\setminus
  N(K))\cup N(K')$: it is clear that every neighbor of $R'$ is in
  $N(R)\cup N(K')$ (as it cannot be in $K'$) and every vertex of
  $N(K)\setminus N(K')$ is fully contained in $K'$. Thus $|N(R')|\le
  |N(R)|$ follows from $|N(K')|\le |N(K)|$. Furthermore, the
  connectivity of $G[K']$ and $K\subset K'$ implies that $K'$ contains
  a vertex of $N(K)\subseteq N(R)$ and therefore $K'\cup N(K')$
  contains a vertex of $R$. This means that $R$ is a proper subset of
  $R'$ with $|N(R')|\le |N(R)|$, contradicting the assumption that $R$
  is a $W$-closest set.
\end{proof}

\subsection{Random sampling of important separators---simplified proof}
\label{sec:rand-select-sets}

In this section, we present a simpler version of the proof of
Theorem~\ref{th:randreduceabstract}, where the probability of
success is double exponentially small in ${\size}$.  This simpler
proof highlights the main idea of the randomized reduction. The full
proof, which improves the probability to $2^{-O({\size}^3)}$ with
additional ideas, appears in Section~\ref{sec:rand-select-sets-full}.

By Lemma~\ref{lem:impsep} we can enumerate every separator of size at most ${\size}$ that is an important $v-W$ separator for some $v$. 
\begin{definition}
  The set $\I_{\size}$ contains a set $S\subseteq V(G)\setminus W$ if $S$ is an
  important $v-W$ separator of size at most ${\size}$ for some vertex
  $v\in V(G)\setminus (W\cup S)$.
 \end{definition}
 By Lemma~\ref{lem:impsep}, the size of $\I_{\size}$ is at most $4^{\size}\cdot |V(G)|$
  and we can construct $\I_{\size}$ in time $O^*(4^{\size})$.

 Recall that the shadow of a set $S$ is the set of vertices not
 reachable from $W$ in $G\setminus S$.  By
 Lemma~\ref{lem:closestboundary}, every vertex of the shadow of $N(R)$
 is covered by the shadow of a member of $\I_{\size}$ that is a subset
 of $N(R)$. This means that the shadow of $2^{{\size}}$ members of
 $\I_{\size}$ fully cover the shadow of $N(R)$.  This suggests that we
 may construct a set $Z$ satisfying the conditions of
 Theorem~\ref{th:randreduceabstract} by guessing these members of $\I_{\size}$ and
 obtaining $Z$ as the union of the shadows of the selected
 sets. However, in general the size of $\I_{\size}$ cannot be bounded
 as a function of ${\size}$ only.
  Thus
 complete enumeration of all possible ways of selecting $2^{{\size}}$ members of $\I_{\size}$ 
 is not feasible. Instead, we randomly select a subset of $\I_{\size}$ and
 hope that it contains these at most $2^{{\size}}$ members and it does
 not contain any member of $\I_{\size}$ whose shadow intersects $N(R)$.

 The probability of randomly selecting a member of $\I_{\size}$ should not be too
 high, because we want to avoid selecting any member whose shadow
 contains a vertex of $N(R)$. We need a bound on the number such members of $\I_{\size}$. Intuitively, the bound of Lemma~\ref{lem:impsep} on the
 number of important separators should imply that each vertex of
 $N(R)$ is contained in the shadow of a bounded number of members of $I_{\size}$, but
 in order to make this claim precise, we need to consider a slightly
 different notion of a shadow:

\begin{definition}
The {\em exact shadow} of a set $S\subseteq V(G)\setminus W$ contains those vertices $v\in V(G)\setminus (W\cup S)$ for which $S$ is a minimal $v-W$ separator.
\end{definition}
For example, in Figure~\ref{fig:shadow}, set $C_2$ is in the exact shadow of $S$, but $C_1$ is not, as a 2-vertex subset of $S$ separates every vertex of $C_1$ from $W$.

The following lemma is true only for exact shadows: the bound in (2)
is not true with the original definition of shadow.
\begin{lemma}\label{lem:impsepexact}
\begin{enumerate}
\item[(1)] For every $S\in \I_{\size}$, we have that $v\in V(G)\setminus (W\cup S)$ is in the exact shadow of $S$ if and only if $S$ is an important $v-W$ separator.
\item[(2)] Each vertex $v\in V(G)\setminus W$ is contained in the exact shadow of at most $4^{{\size}}$ members of $\I_{\size}$.
\end{enumerate}
\end{lemma}
\begin{proof}
  (1) By definition, if $S$ is an important $v-W$ separator, then $S$
  is a minimal $v-W$ separator, hence $v$ is in the exact shadow of
  $S$. For the other direction, suppose that $v$ is in the exact
  shadow of some $S\in \I_{\size}$. By definition of $\I_{\size}$, there is a vertex
  $u\in V(G)\setminus (W\cup S)$ such that $S$ is an important $u-W$
  separator.  If $S$ is not an important $v-W$ separator, then (as
  the definition of exact shadow implies that $S$ is a minimal $v-W$
  separator) there is a $v-W$ separator $S'$ with $|S'|\le |S|$ and
  such that a superset of vertices is reachable from $v$ in
  $G\setminus S'$ compared to $G\setminus S$.

  We claim that $S'$ is a $u-W$ separator as well. Suppose that there
  is a $u-W$ path $P$ in $G\setminus S'$. This path has to go through
  $S\setminus S'$; let $s$ be the first vertex of $S\setminus S'$ on
  $P$ when going from $u$ to $W$. Since $S$ is a minimal $v-W$
  separator, $s$ has a neighbor reachable from $v$ in $G\setminus S$
  and hence in $G\setminus S'$. Therefore, $s\not\in S'$ is also
  reachable from $v$ in $G\setminus S'$. It follows that $s$ is
  reachable from both $u$ and $v$ in $G\setminus S'$, i.e., $u$ and
  $v$ are in the same component of $G\setminus S'$, contradicting the
  assumption that $S'$ is a $v-W$ separator.

  Next we show that every vertex $r$ reachable from $u$ in $G\setminus
  S$ is reachable from $u$ in $G\setminus S'$. Let $P$ be an $u-r$
  path in $G\setminus S$ and suppose that it contains a vertex $q\in
  S'\setminus S$. As $S'$ is a minimal $v-W$ separator, there is a
  $q-W$ path $Q$ that intersects $S'$ only in $q$. The concatenation
  of the prefix of $P$ ending at
  $q$ 
  and $Q$ is a $u-W$ walk, hence $Q$ has to contain a vertex $q'\in
  S$. Vertex $q$ cannot be on $P$; in particular, $q'\neq q$. By the
  definition of $Q$, this vertex $q'$ has to be in $S\setminus S'$ and
  hence it is reachable from $v$ in $G\setminus S'$. However, the
  subpath of $Q$ from $q'$ to $W$ does not contain any vertex of $S'$,
  meaning that $v$ is reachable also from $W$ in $G\setminus S'$, a
  contradiction.  This shows that every vertex reachable from $u$ in
  $G\setminus S$ remains reachable in $G\setminus S'$, contradicting
  the assumption that $S$ is an important $u-W$ separator. Therefore,
  $S$ is indeed an important $v-W$ separator.

  (2) By Lemma~\ref{lem:impsep}, there are at most $4^{\size}$
  important $v-W$ separators of size at most ${\size}$, thus by (1),
  vertex $v$ can be contained in the exact shadows of at most that
  many members of $\I_{\size}$.
\end{proof}

Combining Lemmas~\ref{lem:closestboundary} and \ref{lem:impsepexact}, we immediately have:
\begin{proposition}\label{prop:impsepexact2}
Let $R$ be a $W$-closest set and let $S=N(R)$. Then every vertex $v\not\in R\cup N(R)$ is in the exact shadow of an some $S_v\in \I_{\size}$ with $S_v\subseteq S$.
\end{proposition}
We use Prop.~\ref{prop:impsepexact2} to bound the probability that the
constructed set $Z$ satisfies the second condition of
Theorem~\ref{th:randreduceabstract}. We need the following simple
observation to argue that the selection of these sets does not interfere
with the first condition of Theorem~\ref{th:randreduceabstract}.
\begin{lemma}\label{lem:shadowdisjoint}
  Let $R$ be a $W$-closest set and let $S=N(R)$
  and $S'\subseteq S$. Then the shadow of $S'$ is disjoint from $S$.
\end{lemma}
\begin{proof}
  Suppose that $v\in S$ is in the shadow of $S'\subseteq S$, i.e., $v\not\in S'$
  and $S'$ is a $v-W$ separator.  As $v\in N(R)$, vertex $v$ has a
  neighbor $r\in R$.  We can assume that every component of $G[R]$
  contains a vertex of $W$: otherwise removing a component disjoint
  from $W$ strictly decreases $R$ without increasing $|N(R)|$,
  contradicting the assumption that $R$ is a $W$-closest set.  This
  means that there is a path from $r$ to $W$ fully contained in
  $R$. It follows that there is a path from $v$ to $W$ fully contained
  in $R\cup \{v\}$, which is disjoint from $S'$, contradicting the
  assumption that $v$ is in the shadow of $S'$.
\end{proof}

In the simplified proof of Theorem~\ref{th:randreduceabstract},
we select members of $\I_{\size}$ uniformly at random and take the union of their exact
shadows. In light of Lemmas \ref{lem:closestboundary} and
\ref{lem:impsepexact}, there is a set of at most $2^{\size}$ members of $\I_{\size}$
that have to be selected and there is a set of at most $N(R)\cdot
4^{\size}$ members of $\I_{\size}$ that have to avoided in order for the random selection
to be successful.

\begin{proof}[Simplified proof of Theorem~\ref{th:randreduceabstract}.]
  The algorithm $\randset(G,W,{\size})$ first constructs the set $\I_{\size}$; by Lemma~\ref{lem:impsep}, the size of
  $\I_{\size}$ is $O^*(4^{\size})$ and can be constructed in time $O^*(4^{\size})$.  Let
  $\I_{\size}'$ be the subset of $\I_{\size}$ where each element from
  $\I_{\size}$ occurs with probability $\frac{1}{2}$ independently
  at random. Let $Z$ be the union of the exact shadows of every set in
  $\I_{\size}'$. We claim that the set $Z$ satisfies the requirement
  of the theorem.

  Let $R$ be a $W$-closest set and let $S=N(R)$. Let
  $X_1,X_2,\ldots,X_d\in \I_{\size}$ be the members of $\I_{\size}$ that are fully contained
  in $S$. As $|S|\le {\size}$, we have $d\le 2^{\size}$.  By
  Lemma~\ref{lem:shadowdisjoint}, we have that the exact shadow of $X_j$
  is disjoint from $S$
  for every $j\in[d]$.  Now consider the following events:
\begin{enumerate}
\item[(E1)]$Z\cap S= \emptyset$
\item[(E2)] the exact shadow of $X_{j}$ is a subset
            of $Z$  for every $j\in [d]$.
\end{enumerate}
Note that by Prop.~\ref{prop:impsepexact2}, event (E2) implies that
the shadow of $S$ is fully contained in $Z$, i.e., $V(G)\setminus
(R\cup N(R))\subseteq Z$.  Our goal is to show that with probability
$2^{-2^{O({\size})}}$, events (E1) and (E2) both occur.

Let $A=\{X_1,X_2,\ldots,X_d\}$ and let $B$ contain those sets in $\I_{\size}$
whose exact shadows intersect $S$. By Lemma~\ref{lem:impsepexact},
each vertex of $S$ is contained in the exact shadow of at most $4^{{\size}}$ members of $\I_{\size}$.  Thus $|B|\leq |S|\cdot 4^{{\size}}\leq p\cdot
4^{{\size}}$. If no member of $B$ is selected into $\I_{\size}'$, then event (E1)
occurs. If every member of $A$ is selected $\I_{\size}'$, then event (E2)
occurs. Thus the probability that both (E1) and (E2) occur is bounded
from below by the probability of the event that every element from $A$
is selected and no element from $B$ is selected. Note that $A$ and $B$
are disjoint: $A$ contains only sets whose exact shadows are disjoint
from $S$, while $B$ contains only sets whose exact shadows intersect
$S$.  Therefore, the two events are independent and the probability
that both events occur is at least
\[
\Big(\frac{1}{2}\Big)^{2^{p}}\Big(1-\frac{1}{2}\Big)^{p\cdot 4^{p}} = 2^{-2^{O(p)}}
\]
\end{proof}

\subsection{Random sampling of important separators---full proof}
\label{sec:rand-select-sets-full}

In order to optimize the success probability, we perform the
randomized selection of important separators in two phases: first we select some members of $\I_{\size}$ and
add new edges to the graph and in the second phase we restrict our
attention to members of $\I_{\size}$ that induce cliques in the modified graph.
 We observe that important separators that induce cliques are nested,
hence we can get a bound of ${\size}$ instead of $4^{{\size}}$ for
the number of such separators. 

\begin{lemma}\label{lem:importantclique}
For every vertex $v\in V(G)\setminus W$, there are at most 
${\size}$ important $v-W$ separators of size at most ${\size}$ inducing a clique.
\end{lemma}
\begin{proof}
  Every minimal $v-W$ separator arises as $N(X)$ for some set $X$ with
  $v\in X$ and $G[X]$ connected. The bound follows from observing that
  important separators inducing cliques are nested.  That is, we show
  that if $X_1$ and $X_2$ are connected sets containing $v$ such that
  $N(X_1)$ and $N(X_2)$ are important $v-W$ separators inducing
  cliques, then either $X_1\subset X_2$ or $X_2\subset X_1$.

  Suppose that $X_1\setminus X_2$ and $X_2\setminus X_1$ are both
  nonempty.  If $X_1\setminus X_2\neq\emptyset$ and $X_1$ is
  connected, then there is a vertex $x_1\in X_1\cap N(X_2)$. As
  $N(X_2)$ is a clique, every vertex of $N(X_2)$ is adjacent with
  $x_1$, implying that $N(X_2)\subseteq X_1\cup N(X_1)$. If
  $X_2\setminus X_1\neq\emptyset$, then a symmetrical argument shows
  that $N(X_1)\subseteq X_2\cup N(X_2)$.  We claim that $N(X_1\cup
  X_2)\subseteq N(X_1)\cap N(X_2)$ and hence $|N(X_1\cup X_2)|\le
  |N(X_1)|,|N(X_2)|$; as $X_1\cup X_2\supset X_1,X_2$, this
  would contradict the assumption that $X_1$ and $X_2$ are important
  components. Consider a vertex $u\in N(X_1\cup X_2)$, which must have
  a neighbor $w\in X_1\cup X_2$.  If $w \in X_1 \cap X_2$, then $u \in
  N(X_1) \cap N(X_2)$ and we are done. Suppose without loss of
  generality that $w\in X_1 \setminus X_2$. Then $u\in N(X_1)\subseteq
  X_2\cup N(X_2)$, but $u\not\in X_2$ by definition, hence $u$ has to
  be in $N(X_2)$ as well.

  Suppose now that $X_1$, $X_2$, $\dots$, $X_t$ are connected sets containing
  $v$ such that $N(X_1)$, $N(X_2)$, $\dots$, $N(X_t)$ are important $v-W$
  separators inducing cliques.  We have shown that the $X_i$'s form a
  chain, i.e., we can assume without loss of generality that
  $X_1\subset X_2 \subset \dots \subset X_t$. This means that there are at most
  ${\size}$ of them, as the definition of important separator implies
  that $|N(X_1)|<|N(X_2)|<\dots< |N(X_t)|$ has to hold.
\end{proof}
By Lemma~\ref{lem:impsepexact}(1), we have the following the bound:
\begin{lemma}\label{lem:importantclique2}
Every vertex $v\in V(G)\setminus W$ is contained in the exact shadow of at most 
${\size}$ sets $X\in \I_{\size}$ such that $G[N(X)]$ is a clique.
\end{lemma}

\begin{proof}[Full proof of Theorem~\ref{th:randreduceabstract}.]
  The randomized algorithm consists of two phases. For consistency of
  notation, let $G_1=G$ and $\I_{\size,1}=\I_{\size}$. In the first phase, we select a subset of
  $\I_{\size}$ and obtain $G_2$ for $G_1$ by making the selected sets
  cliques.
Let $\I_{\size,2}$ be defined as $\I_{\size,1}$, but for graph $G_2$: 
$S$ is in $\I_{\size,2}$ if it  is an
  important $v-W$ separator of size at most ${\size}$ for some vertex
  $v\in V(G_2)\setminus (W\cup S)$ in $G_2$.
 In the second phase, we select a subset of $\I_{\size}$ inducing
  cliques in $G_2$ and obtain $Z$ as the union of the exact shadows of
  the selected sets.

  \textbf{Phase 1.} In the first phase, we select a subset
  $\I_{\size,1}'\subseteq \I_{\size,1}$ by putting every set of $\I_{\size,1}$ into $\I_{\size,1}'$ with
  probability $p_1=4^{-{\size}}$ independently at random. Then we make
  every set $X\in \I_{\size,1}'$ a clique; let $G_2$ be the graph obtained this
  way.  

  Let $R$ be a $W$-closest set and let $S=N(R)$.  By
  Proposition~\ref{prop:impsepexact2}, there exists a subcollection $A_2$ of $\I_{\size,1}$, all being subsets of $S$, such that $V(G)\setminus (R\cup S)$ is
  covered by the exact shadows of the sets in $A_2$.  Let us estimate the probability that the
  events
\begin{enumerate}
\item[(E1)] Every $S'\in A_2$ induces a clique in $G_2$.
\item[(E2)] Every $S'\in A_2$ has the same exact shadow in $G_1$ and in $G_2$.
\item[(E3)] Every $S'\in A_2$ is in $\I_{\size,2}$.
\end{enumerate}
occur.

Let us make a subset $A_1$ of $A_2$ such that for every $S_2\in A_2$ and $x,y\in S_2$, there is
a set $S_1\in A_1$ with $x,y\in S_1$. In other words, the sets in
$A_1$ cover every pair $\{x,y\}$ of vertices covered by the sets in
$A_2$.  Since there are $\binom{|S|}{2}\le \binom{{\size}}{2}$ such
pairs, it is clear that there exists a collection $A_1$ of size at
most $\binom{{\size}}{2}$.  Observe that, by
Lemma~\ref{lem:shadowdisjoint}, the shadow of every set in $A_1$ is disjoint from
$S$.

Let $B_1$ contain those members of $\I_{\size,1}$ whose exact shadows intersect $S$; by
Lemma~\ref{lem:impsepexact}, we have $|B_1|\le |S|\cdot 4^{\size} \le
{\size} \cdot 4^{\size}$. We claim that if every member of $A_1$ is in
$\I_{\size,1}'$ and no member of $B_1$ is in $\I_{\size,1}'$, then (E1--E3) occur.

Consider an $S'\in A_2$. Assuming that
every member of $A_1$ is in $\I_{\size,1}'$, the set $G_2[S']$ becomes a
clique. This shows (E1).

To show (E2), that is, that  $S'\in A_2$ has the same exact shadows in $G_1$ and $G_2$, we
show that a subset $S'\subseteq S$ is a $v-W$ separator for some
vertex $v$ in $G_1$ if and only if it is in $G_2$. This shows that
$S'$ is a minimal $v-W$ separator in $G_1$ if and only if it is in
$G_2$, implying the equalities of the exact shadows. One direction is
clear, as $G_1$ is a subset of $G_2$. For the other direction, suppose
that $S'$ is not a $v-W$ separator in $G_2$. Let $K$ be the connected
component of $v$ in $G_1\setminus S'$;  by
assumption $K$ is disjoint from $W$. Then there have to be two
vertices $a\in K$ and $b\not\in K\cup S'$ that are adjacent in $G_2$
but not in $G_1$. The reason why $a$ and $b$ are adjacent in $G_2$ is
that there is some $X\in \I_{\size,1}'$ with $a,b\in X$. As we assumed that no
member of $B_1$ is in $\I_{\size,1}'$, this means that the exact shadow of $X$
is disjoint from $S$ (and hence from $S'$). As $X\in \I_{\size,1}$, it is an important (hence minimal) $q-W$ separator for some vertex $q$ in its exact shadow. This means that there are paths from $q$ to $a$ and $b$ in the exact shadow of $X$.
Therefore, there is a path
$P$ from $a$ to $b$ in $G_1$ whose internal vertices are in the exact shadow of $X$, hence disjoint from
$S'$. It follows that $b$ is also in the component $K$ of $v$ in
$G_1\setminus S'$, a contradiction.

Finally, let us show (E3).  As $S'\in \I_{\size,1}$, it is an important $v-W$ separator for
some  vertex $v$.  Again, let $K$ be the connected component of $v$ in
$G_1\setminus S'$. By the previous paragraph, $S'$ is a $K-W$
separator in $G_2$. This implies that $S'$ is an important $v-W$
separator in $G_2$ as well: if there is a separator $S''$
contradicting that $S'$ is an important $v-W$ separator in $G_2$, then
$S''$ is a $v-W$ separator in $G_1$ as well (as $G_1$ is a subgraph of
$G_2$) and at least one vertex of $S'$ is reachable from $v$ in
$G_1\setminus S''$, which means that $S''$ contradicts that $S'$ is an
important $v-W$ separator in $G_1$.

We can conclude that the probability that (E1--E3) occur can be
bounded from below by the probability of the event that every set in
$A_1$ is selected and no set from $B_1$ is selected. As the sets $A_1$
and $B_1$ are disjoint (recall that the exact shadow of every member
of $A_1$ is disjoint from $S$ by Lemma~\ref{lem:shadowdisjoint} while
the exact shadow of every member of $B_1$ intersects $S$ by
definition), this probability is at least 
\[(1-4^{-{\size}})^{{\size}\cdot 4^{\size}}\cdot (4^{-{\size}})^{{\size}^2}\ge e^{-2{\size}}\cdot 4^{-{\size}^3}=2^{-O({\size}^3)}
\]
(in the inequality, we use that $1+x\ge
\exp(x/(1+x))$ for every $x>-1$ and $1-4^{-{\size}}\ge 1/2$).

\textbf{Phase 2.} $\I_{\size,2}'$ be
a subset of $\I_{\size,2}$ where every $X\in\I_{\size,2}$ with $G_2[X]$ being a clique appears with probability
$p_2=1-2^{-{\size}}$ independently at random (and if a set $X\in \I_{\size,2}$ does not induce a clique in $G_2$, then it is never selected). Let $Z$ be the union of
the exact shadows of the sets in $\I_{\size,2}'$. 

If (E1--E3) occur, then every set in $A_2$ is in $\I_{\size,2}$ and they induce cliques in $G_2$. If additionally the events
\begin{enumerate}
\item[(E4)]
$Z\cap S=\emptyset$, and 
\item[(E5)] $A_2\subseteq \I_{\size,2}'$
\end{enumerate}
occur, then every $v\not\in R\cup N(R)$ is in the exact shadow of some $S'\in \I_{\size,2}'$ and $v\in Z$ follows.

Let us estimate the probability that both (E4) and (E5) hold on
condition that (E1--E3) hold. Let $B_2$ contain those members of
$\I_{\size,2}$ (inducing cliques) whose exact shadow in $G_2$ intersects $S$; we have $|B_2|\le
{\size}|S|\le {\size}^2$ (by Lemma~\ref{lem:importantclique2}, every
vertex of $S$ is contained in the exact shadow of at most ${\size}$
members of $\I_{\size,2}$ inducing cliques in $G_2$).  If no member of $B_2$ is selected, then no
exact shadow of a set of $\I_{\size,2}'$ contains a vertex of $S$, and hence
$Z\cap S=\emptyset$.  Note that $A_2$ and $B_2$ are disjoint: by (E2),
every $S'\in A_2$ has the same exact shadow in $G_1$ and $G_2$,
therefore the exact shadow of $S'\in A_2$ is disjoint from $S$ in
$G_2$ as well.  Therefore, the probability that (E4) and (E5) hold
can be bounded from below by the probability of the event that every
member of $A_2$ is selected and no member of $B_2$ is selected, which
is at least
\[
(2^{-{\size}})^{{\size}^2}\cdot (1-2^{-{\size}})^{2^{\size}}\ge
2^{-{\size}^3}\cdot e^{-2} 2^{-O({\size}^3)}
\]
(again,  we use that $1+x\ge
\exp(x/(1+x))$ for every $x>-1$ and $1-2^{-{\size}}\ge 1/2$).

Taking into account the probability of success in both phases, we get
that for each $W$-closest set $R$, the set $Z$ satisfies the requirements with probability
$2^{-O({\size}^3)}$.
\end{proof}

\subsection{Derandomization}
\label{sec:derandomization}By running $2^{O({\size}^3)}$ times the algorithm of Lemma~\ref{lem:shadowlessrand},  we get
a collection of instances such that at least one of them satisfies the requirements of Lemma~\ref{lem:shadowless}
with arbitrary large constant probability. 
To obtain a deterministic version of 
Lemma~\ref{lem:shadowless}, we  derandomize the algorithm of
Theorem~\ref{th:randreduceabstract} using the standard technique of splitters.

\begin{lemma}\label{lem:derandreduceabstract}
  There is an algorithm $\derandset(G,W,{\size})$ that, given a graph
  $G$, a set $W\subseteq V(G)$, and an integer ${\size}$,  produces
  $t=2^{O({\size}^3)}\log^2 |V(G)|$ subsets $Z_1$, $\dots$, $Z_t$ of
  $V(G)\setminus W$ such that the following holds.
 For
  every closest set $R$ with $|N(R)|\le {\size}$, there is at least one $1\le i \le t$ with
\begin{enumerate}
\item $N(R)\cap Z_i=\emptyset$, and
\item $V(G)\setminus (R\cup N(R))\subseteq Z_i$.
\end{enumerate}
\end{lemma}
\begin{proof}
An {\em $(n,r,r^2)$-splitter} is a family of functions from
$[n]$ to $[r^2]$ such that for any subset $X\subseteq [n]$ with
$|X|=r$, one of the functions in the family is injective on $X$.
Naor, Schulman, and Srinivasan \cite{NaorSS95} gave an explicit construction of an
$(n,r,r^2)$-splitter of size $O(r^6\log r \log n)$. 

Observe that in the first phase of the algorithm of
Theorem~\ref{th:randreduceabstract}, a random subset of a universe $\I_{\size,1}$ of size
$n_1=|\I_{\size,1}|\le 4^{\size}\cdot n$ is selected, where $n=|V(G)|$. There is a collection
$A_1\subseteq \I_{\size,1}$ of $a_1\le {\size}^2$ sets and a collection $B_1\subseteq
\I_{\size,1}$ of $b_1 \le {\size}\cdot 4^{\size}$ sets such that if every set in $A_1$ is
selected and no set in $B_1$ is selected, then (E1--E3) hold.
Instead of selecting a random subset, we try every function $f$ in
an $(n_1,a_1+b_1,(a_1+b_1)^2)$-splitter family and every subset $F\subseteq
[(a_1+b_1)^2]$ of size $a_1$ (there are
$\binom{(a_1+b_1)^2}{a_1}=2^{O({\size}^3)})$ such sets $F$). For a particular choice of $f$ and $F$,
we select those sets $X\in \I_{\size,1}$ for which $f(X)\in F$. By the
definition of the splitter, there will be a function $f$ that is
injective on $A_1\cup B_1$, and there is a subset $F$ such that
$f(X)\in F$ for every $A_1$ and $f(X)\not\in F$ for every $B_1$. For
such an $f$ and $F$, the selection will ensure that (E1--E3) hold.

In the second phase, we select a random subset of universe $\I_{\size,2}$ of
size $n_2\le {\size}n$, and there is a collection $A_2\subseteq \I_{\size,2}$ of size
$a_2\le 2^{\size}$ and a collection $B_2\subseteq \I_{\size,2}$ of size $b_2\le
{\size}^2$ such that if every set in $A_2$ is selected and no set in
$B_2$ is selected, then (E4) and (E5) hold. As in the first phase, we
can replace this random choice by enumerating the functions of an
$(n_2,a_2+b_2,(a_2+b_2)^2)$-splitter and every subset $\overline F\subseteq
[(a_2+b_2)^2]$ of size $b_2$ (there are
$\binom{(a_2+b_2)^2}{b_2}=2^{O({\size}^3)}$ such sets $\overline F$). This
time, we select a set $X\in \I_{\size,2}$ if $f(X)$ is {\em not} in $\overline F$
and it is clear that there is an $f$ and $\overline F$ for which (E4) and
(E5) hold.

Let us bound the number of branches of the algorithm. In both phases,
the size of the splitter family is $2^{O({\size})}\cdot \log n$ and
the there are $2^{O({\size}^3)}$ possible $F$. (Note that
the splitter family can be constructed in time polynomial in the size
of the family.) Thus the algorithm produces
$2^{O({\size}^3)}\cdot \log^2 n$ sets.
\end{proof}

\section{Reduction to the bipedal case}
\label{sec:reduct-biped-case}

Let $(G,{\bf T},W,{\size})$ be an instance of the \mccs\ problem.  Let
us call a component of $G \setminus W$ having at least two legs a
\emph{non-trivial component} of $G$ w.r.t. $W$ (when the context is
clear, we will just refer to a non-trivial component).  As the
solution of \mccs\ has to be a set $S$ that is disjoint from $W$ and a
multiway cut of $W$, the number of non-trivial components is a lower
bound on the size of the solution.

We present an algorithm that either solves the given instance of the \mccs\
problem or produces a set of instances of the \bmcc\ problem whose number is
bounded by a function of ${\size}$ and such that if the considered instance of
the \mccs\ problem has a shadowless solution then one of the output
instances of the \bmcc\ problem has a solution. In addition, any (not necessarily
shadowless) solution of any of these output instances is a solution of the
input instance of the \mccs\ problem. The key ingredient of this algorithm is
a procedure that, given an instance of the \mccs\ problem where at least
one component has more than $2$ legs, reduces this instance to a  set of instances 
whose number is bounded by a function of ${\size}$ and
such that in each instance either the parameter is decreased or the
number of non-trivial components is increased.

The main idea for the branching is the following. Let $B$ be a set of
vertices in $G\setminus W$ and let $S$ be a hypothetical shadowless solution for
\mccs. We try to guess what happens to each vertex of $B$ in the
solution $S$. It is possible that a vertex $v\in B$ is in $S$; in this
case, we delete $v$ from the instance and reduce the parameter. 
Otherwise, as the solution is shadowless, $v$ has to be in the
same component as precisely one $w\in W$ (since $S$ is a multiway cut of
$W$).  In this case, identifying $v$ and $w$ does not change the
solution.

The following lemma formalizes these observations.  Given a set $B$ of
vertices in $G\setminus W$ and a function $f:B \rightarrow W$, we
denote by $G_f$ the graph obtained by replacing each set $\{w\} \cup
f^{-1}(w)$ with a single vertex (with removal of loops and multiple
occurrences of edges). To simplify the presentation, we will assume
that this new vertex is also named $w$.  We denote by ${\bf T}_f$ the
set of terminal pairs where each vertex $v\in B$ is replaced by
$f(v)$, and we denote by ${\bf T}\setminus v$ the set where every pair
involving the vertex  $v$ is removed.

\begin{lemma}\label{branchcases}
Let $K$ be a non-trivial component of $G\setminus W$ with set of legs $\wleg$ and let $B \subseteq K$.
If $(G,{\bf T},W,{\size})$ has a shadowless solution, then one of the following 
statements is true.
\begin{itemize}
\item There is a $v \in B$ such that the instance $(G \setminus v,{\bf
    T}\setminus v,W,{\size}-1)$ has a shadowless solution.
\item There is a function $f:B \rightarrow \wleg$ such that instance $(G_f,{\bf T}_f,W,{\size})$ has
a shadowless solution.
\end{itemize}
Moreover, if any of the above instances has a solution, then $(G,{\bf T},W,{\size})$ has a solution as well.
\end{lemma}
\begin{proof}
Assume that $(G,{\bf T},W,{\size})$ has a
shadowless solution $S$. Then it either intersects or does not intersect with
$B$. In the former case, we can specify a $v \in S \cap B$ such that
$S \setminus \{v\}$ is a shadowless 
solution of $(G \setminus v,{\bf T}\setminus v,W,{\size}-1)$.
In the latter case, we can assign each $v \in B$
\emph{precisely} one vertex $f(v)$ of $\wleg$ such that vertex $v$
belongs to the same component of $G\setminus S$ as $f(v)$.  It is not
hard to see that $S$ is a shadowless solution of
$(G_f,{\bf T}_f,W,{\size})$. 

For the second statement, we observe that the existence of a solution for any
of the above instances implies the existence of a solution for $(G,{\bf T},W,{\size})$. 
This is certainly true in the first case, where we
delete a vertex and decrease the parameter by $1$. In the second
case, the statement follows from the fact that replacing $G$ with
$G_f$ by identifying vertices cannot make the problem any easier.
\end{proof}
Lemma \ref{branchcases} determines a set of recursive calls to
be applied in order to find a solution for the given instance $(G,{\bf
  T},W,{\size})$ of the \mccs\ problem, if a shadowless solution is guaranteed to exist. It is clear that in each step,
the number of directions we branch into is bounded by a function of
${\size}$, $|B|$, and $|W|$ (observe that the number of  functions $f:B\to \wleg$ can be bounded by
$|\wleg|^{|B|}\le |W|^{|B|}$). However, in order to ensure that the size of the
search tree is bounded, we need to ensure that the {\em height} of the
search tree is bounded as well. This is obvious for the first type
of branches, as ${\size}$ decreases. The following property ensures
that in every branch of the second type, either the number of
nontrivial components increases or we get an instance that trivially
has no solution.

\begin{definition}\label{def:shattering}
Let $K$ be a non-trivial component and let $\wleg \subseteq W$ be its set of legs.
Let $B$ be a subset of $K$.
We say that $B$ is a \emph{shattering set} if for any function $f:B \rightarrow \wleg$
one of the following statements is true regarding the instance $(G_f,{\bf T}_f,W,{\size})$ of the \mccs.
\begin{itemize}
\item There is a $w\in \wleg$ such that there is no $w-(\wleg\setminus \{w\})$
  separator of size  at most ${\size}$ 
  in $G_f[K \cup \wleg]$.
\item  The number of non-trivial components of $G_f \setminus W$ is greater than the number of non-trivial
components of $G \setminus W$. 
\end{itemize}
\end{definition}
Note that the first possibility includes the case when $G_f[\wleg]$ is
not an independent set (recall that an $X-Y$ separator is disjoint
from $X\cup Y$ by definition).   In
Section~\ref{sec:finding-shattering-set}, we present a polynomial-time
algorithm for finding a shattering set.
\begin{lemma} \label{lem:findshattering}
Given an instance $(G,{\bf T},W,{\size})$ of the \mccs\ problem
and a component $K$ of $G\setminus W$ with more than two legs, we can
find a shattering set $B\subseteq K$ of size at most $3{\size}$ in polynomial time.
\end{lemma}
With Lemma~\ref{lem:findshattering} in mind, we are ready to prove Lemma~\ref{lem:shatterit}, the main statement of this section. 

\begin{proof}[Proof (of Lemma~\ref{lem:shatterit})]
The desired algorithm looks as follows. If the given instance $(G,{\bf T},W,{\size})$
of \mccs\ satisfies one of the following cases, then we can determine the answer
without any further branching:
\begin{itemize}
\item All the terminal pairs of ${\bf T}$ are separated: solve the {\sc multiway cut} problem $(G,W,{\size})$.
\item The parameter is zero while there are unseparated terminals: this is a ``NO'' instance.
\item There is a $w\in W$ such that there is no $w-(W\setminus \{w\})$ separator of
  size at most ${\size}$ in $G$: this is a ``NO'' instance. The situation
  where $W$ is not an independent set is a special subcase of this case.
\item The number of non-trivial components is greater than ${\size}$: this is a ``NO'' instance since
each non-trivial component contributes at least one vertex to any solution.
\item Every component has at most two legs: this is an instance of
  \bmcc\ problem and hence it is returned as the output.
\end{itemize}
Otherwise, we choose a component $K$ of $G \setminus W$ having more than
two legs, and use Lemma~\ref{lem:findshattering} to compute a shattering subset $B$ of $K$ of size at most
$3{\size}$. We apply recursively the branches
specified in the statement of Lemma \ref{branchcases}. If the ``YES''
answer is obtained on at least one of these branches, then we return
``YES''. If all the branches return ``NO'', we return ``NO''.  According to Lemma \ref{branchcases},
the resulting answer is correct. Furthermore, assume that no one of branches
produces a ``YES'' or ``NO'' answer. Then, according to Lemma \ref{branchcases},
if the parent instance has a shadowless solution, then the instance on
one of the branches has a shadowless solution. It is also not hard to notice that
any solution for a branch instance can be easily transformed into a solution of the
parent instance. Applying this argument inductively, we conclude that the same
relationship exists between the original instance $(G,{\bf T},W,{\size})$ and the
\bmcc\ problem instances at the leaves of the recursion tree. 

To bound the number of leaves of the recursion tree, let us define $\kappa$ to be
the number of nontrivial components. Observe that removing a vertex of
$V(G)\setminus W$ from $G$ can decrease the number of nontrivial
components only by at most one. Thus inspection of
Lemma~\ref{branchcases} shows that the measure $2{\size}-\kappa$ strictly
decreases in each branch. This means that the height of the search
tree is at most $2{\size}$. The number of branches in each step can be bounded by
$3{\size}+|W|^{3{\size}}$. Thus the
number of leaves of the recursion tree can be generously bounded by
$2^{O(({\size}+\log|W|)^3}$. Taking into account that the runtime \emph{per node}
of the recursion tree is polynomial, it follows that the runtime of this algorithm
is $O^*(2^{O(({\size}+\log|W|)^3)})$.
\end{proof}

\subsection{Finding a shattering set}\label{sec:finding-shattering-set}
 We try to find a shattering set by selecting a set that separates
one leg from all the other. If it is {\em not} a shattering set, then we can
characterize quite well how it can look like, and where should we
continue our search for a shattering set. Let us start with two simple
lemmas.
\begin{lemma}\label{uniquemwaycut}
  Let $K$ be a non-trivial component with a set $\wleg$ of at least 3 legs. If $G[M_1]$ and $G[M_2]$ are both connected for
  two disjoint sets $M_1,M_2\subseteq K$, then at most one of $M_1$ and $M_2$ can
  be a multiway cut of ($G[K \cup \wleg],\wleg)$.
\end{lemma}
\begin{proof}
Assume the opposite.  Since no two vertices of $\wleg$ belong to the same component of $G[K
  \cup \wleg] \setminus M_1$ and $|\wleg| \geq 3$, we can specify two
  vertices $w'$ and $w''$ of $\wleg$ whose respective components $C'$ and
  $C''$ in $G[K \cup \wleg] \setminus M_1$ are disjoint from the
  connected set $M_2$. As $G[K]$ is connected, there is a $w'-w''$ path in $G[K\cup \wleg]$
  that first uses vertices from $C'$, then vertices from (the connected set)
  $M_1$, then vertices from $C''$. This path is disjoint from $M_2$,
  contradicting the assumption that $M_2$ is a multiway cut.
\end{proof}

\begin{lemma} \label{GraceCriterion}
  Let $K$ be a non-trivial component with a set $\wleg$ of at least 3 legs.
Let $B\subseteq K$ be a non-shattering set.
Then there is exactly one connected component of
$G[K \setminus B]$ that is a multiway cut of
($G[K \cup \wleg],\wleg)$.
\end{lemma}

\begin{proof}
  Let $f: B \rightarrow \wleg$ be the mapping witnessing that $B$ is
  not a shattering set. Let $K' \subseteq K\setminus B$ be the unique
  non-trivial component of $G_f \setminus W$ that is a subset of
  $K$ (witnessing $B$ being a non-shattering set). 
As every neighbor of $K'$ is in $B\cup \wleg$, it is easy to see that $K'$ is a component of $G[K\setminus B]$ as well.
  Furthermore, we claim that $K'$ is a multiway cut of $(G[K \cup
  \wleg],\wleg)$. Otherwise, a path between vertices of $\wleg$ in
  $G[K \cup \wleg]\setminus K'$ would correspond to a walk of $G_f$
  between the same vertices which belong to a non-trivial component
  that is a subset of $K$ but different from $K'$, in contradiction to
  the definition of $f$. Finally, Lemma~\ref{uniquemwaycut} implies that
  $K'$ is the unique connected component of $G[K \setminus B]$ being a
  multiway cut of $G[K \cup \wleg]$.
\end{proof}

\begin{figure}
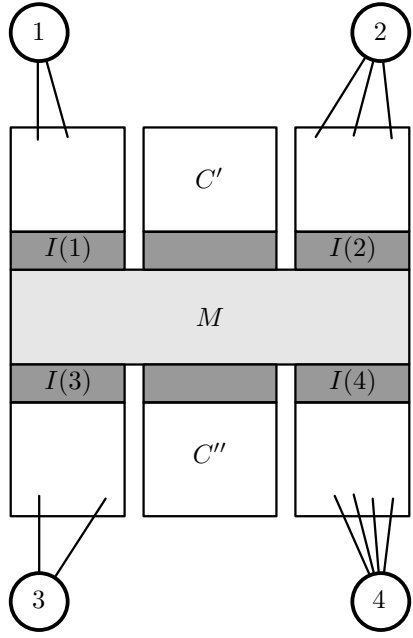

\begin{center}
{\small \svg{0.32\linewidth}{image}}
\caption{$M$ is a multiway cut of the 4 legs $\{1,2,3,4\}$. The dark
region represents the boundary of $M$. Observe that $I(\{1,2,3,4\})$
is a proper subset of the boundary: vertices of the boundary that are
adjacent only to $C'$ and $C''$ are not in $I(w)$ for any $w\in
\{1,2,3,4\}$.} \label{fig:image}
\end{center}
\end{figure}

Let $K$ be a non-trivial component with a set of legs $\wleg$.  Let $M
\subseteq K$ be a multiway cut of $(G[K \cup \wleg],\wleg)$.  We call
$N(M)$ (i.e., the open neighborhood of $M$) the \emph{boundary} of $M$
(which possibly includes vertices of $\wleg$). For each $w \in \wleg$,
the \emph{image} $I(w)$ of $w$ is the set of vertices of $N(M)$
reachable from $w$ in $G[K \cup \wleg] \setminus M$ (the image may
include vertex $w$ itself, but it cannot include any other member of $W$), see Figure \ref{fig:image}. Note that $I(w)$ is nonempty for any $w\in
\wleg$: consider the first vertex of $N(M)$ on a path from $w$ to some
other leg in $\wleg$.  Furthermore, as $M$ is a multiway cut, the sets
$I(w')$ and $I(w'')$ are disjoint for $w'\neq w''$: otherwise, there
would be a $w'-w''$ path disjoint from $M$.  For $X\subseteq \wleg$,
we let $I(X)=\bigcup_{w\in X}I(w)$.  Let us select a distinguished leg
$w^*\in \wleg$. We say that $M$ is \emph{good} if all of the following
conditions are true.

\begin{itemize}
\item $G[M]$ is connected,
\item $N(M)=I(\wleg)$ or, in other words, each vertex of $N(M)$ is
  reachable in $G[K \cup \wleg] \setminus M$ from some vertex of $\wleg$, and
\item $|I(w^*)\setminus \wleg|\le p$ and $|I(\wleg\setminus \{w^*\})\setminus \wleg| \le p$ holds (and hence we have $|N(M)\setminus \wleg|\le 2p$).
\end{itemize}

Our goal is to obtain a shattering set from the boundary of a good
multiway cut. The following lemma gives a polynomial-time algorithm
that either produces a shattering set, or finds a smaller good
multiway cut. Interestingly, the algorithm does not check that the
returned set $B$ is a shattering using Definition~\ref{def:shattering}
directly: this would require trying every function $f:B\to
\wleg$. Instead, the way the set $B$ is produced guarantees that $B$ is
indeed a shattering set.
\begin{lemma} \label{IterationStep}
  Let $K$ be a non-trivial component with a set $\wleg$ of at least 3 legs and a distinguished leg $w^*$.
Let $M$ be a good multiway cut of $(G[K \cup \wleg],\wleg)$.
Then there is a polynomial-time algorithm that either returns
a shattering set of size at most $3p$ or a good multiway cut
$M' \subset M$.
\end{lemma}

\begin{proof}
  The desired algorithm first computes a smallest $I(w^*)-I(\wleg \setminus
  \{w^*\})$ separator $S$ of $G[N(M) \cup M]$ (recall that the images
  are nonempty). Observe that $S$ is an inclusionwise minimal
  $w^*-\wleg\setminus\{w^*\}$ separator in $G[K\cup \wleg]$ (and hence
  nonempty).  We consider three cases:
\begin{enumerate}
\item If $|S|>p$, then the algorithm returns $B:=N(M) \setminus \wleg$
  reporting it as a shattering set.
\item If $|S| \leq p$ and there is a unique connected component $M'$
  of $G[K \setminus (N(M)\cup S)]$ that is a multiway cut of $(G[K \cup
  \wleg],\wleg)$, then the algorithm returns $M'$ reporting it as a good
  multiway cut.
\item If $|S|\leq p$ and there is no such unique $M'$, then the algorithm
returns $B:=(N(M) \cup S) \setminus \wleg$ reporting it as a shattering set.
\end{enumerate}
This algorithm clearly takes polynomial time. The remaining proof
establishes correctness of the algorithm in each of these three cases.

\textbf{Case 1.}  The definition of good multiway cut implies that that $B:=N(M)\setminus \wleg$ has size at most $2p$. We prove that $B$ is a
shattering set. Otherwise, let $f:B\to \wleg$ be a function witnessing
that $B$ it is not a shattering set.  It is not hard to see that $M$
is a connected component in $G_f \setminus W$ whose set of legs is a
subset of $\wleg$. We consider three subcases and arrive to a
contradiction in each of them (see Figure~\ref{fig:case1}).
\begin{figure}
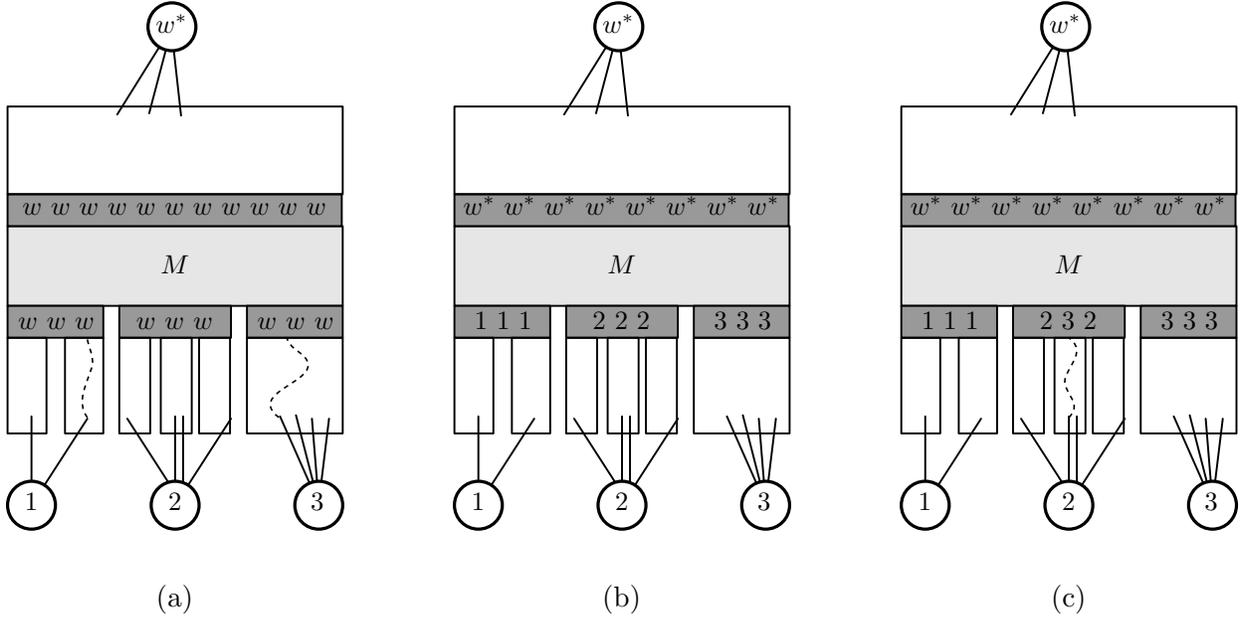

\begin{center}
{\small \svg{0.98\linewidth}{case1}}
\caption{The 3 subcases of Case~1 in Lemma~\ref{IterationStep} for a component with legs $\{w^*,1,2,3\}$. Case~1a: $w$ is the only leg of $M$ in $G_f\setminus W$. The figure shows two paths in two distinct components connecting $w$ to another leg (assuming $w\not\in \{1,3\}$). Case~1b: $f(v)=w$ for every $v\in I(w)$. Case~1c: $M$ is a nontrivial component in $G_f\setminus W$ and $f(v)=3$ for some $v\in I(2)$; the figure shows a $2-3$ path.} \label{fig:case1}
\end{center}
\end{figure}

\textbf{Case 1a.} $M$ is a trivial component of $G_f\setminus W$. 
Let $w$ be the only leg of $M$. Let $w_1$ and $w_2$ be other two distinct legs of $K$ in $G$
that are different from $w$. It follows that $f$ maps every vertex of $I(w_1)\cup I(w_2)$ to $w$
implying that there is a $w-w_1$ and a $w-w_2$ path in $G_f$ whose internal vertices belong to
two different components adjacent to $w_1$ and $w_2$ in $G[K \cup \wleg] \setminus M$. Thus $G_f$ has at
least two non-trivial components that are subsets of $K$, in contradiction to the choice of $f$.

\textbf{Case 1b.} $M$ is a nontrivial component of $G_f\setminus W$
and $f(v)=w$ for every $v\in I(w)$ and $w\in \wleg$ (i.e., each vertex
on the boundary is mapped to its preimage).  As the smallest
$I(w^*)-I(\wleg\setminus \{w^*\})$ separator in $G[N(M)\cup M]$ is
larger than $p$, $G[M \cup \wleg]$ does not have a $w^*-\wleg
\setminus \{w^*\}$ separator of size at most $p$, in contradiction to
$f$ being a witnessing function.

\textbf{Case 1c.} $M$ is a nontrivial component of $G_f\setminus W$
and there are distinct $w_1,w_2 \in \wleg$ such that $f(v)=w_2$ for
some $v\in I(w_1)$. By definition of $I(w_1)$, there is a $w_1-v$ path
in $G$ whose internal vertices are fully contained in $K\setminus
M$. Therefore, there is a $w_1-w_2$ path in $G_f$ whose internal
vertices are disjoint from $M$, implying that $G_f$ has a nontrivial
component that is a subset of $K$, but distinct from the nontrivial
component $M$.  Thus the number of nontrivial components increases, a
contradiction.

\textbf{Case 2.} We show that $M'\subset M$ and $M'$ is a good
multiway cut in this case.  Let us show $M'\subset M$ first.  Clearly,
$M'\neq M$, as $M'$ is disjoint from the (nonempty) set $S\subseteq
M$. Thus $M'\not\subset M$ is only possible if $M'$ is disjoint from
$M$, but Lemma~\ref{uniquemwaycut} implies that the two disjoint
connected sets $M$ and $M'$ cannot be both multiway cuts.

For clarity, from now on we use $I_M(w)$ and $I_{M'}(w)$ for the image
of $w$ on the boundary of $M$ and $M'$, respectively.  Observe that
$I_M(w)\cap N(M')\subseteq I_{M'}(w)$ for every $w\in \wleg$: for
every $v\in I_M(w)\cap N(M')$, there is a $w-v$ path disjoint from
$M$, which is obviously disjoint from $M'\subset M$ as well, and then
$v\in N(M')$ implies $v\in I_{M'}(w)$. We claim that either $I_M(w^*)$
or $I_M(\wleg\setminus \{w^*\})$ is disjoint from $N(M')$. Suppose
that there are two vertices $v_1\in I_M(w^*)\cap N(M')$ and $v_2\in
I_M(\wleg\setminus\{w^*\})\cap N(M')$. Vertices $v_1$ and $v_2$ can be
connected by a path $P$ whose internal vertices are in $M'$ (hence
disjoint from $S$), contradicting the fact that $S$ is an $I_M(w^*)-
I_M(\wleg\setminus\{w^*\})$ separator.  Therefore, either
$N(M')\subseteq I_M(w^*)\cup S$ or $N(M')\subseteq I_M(\wleg\setminus
\{w^*\})\cup S$ holds.
The two possibilities are demonstrated in Figure~\ref{fig:goodmway}.
\begin{figure}
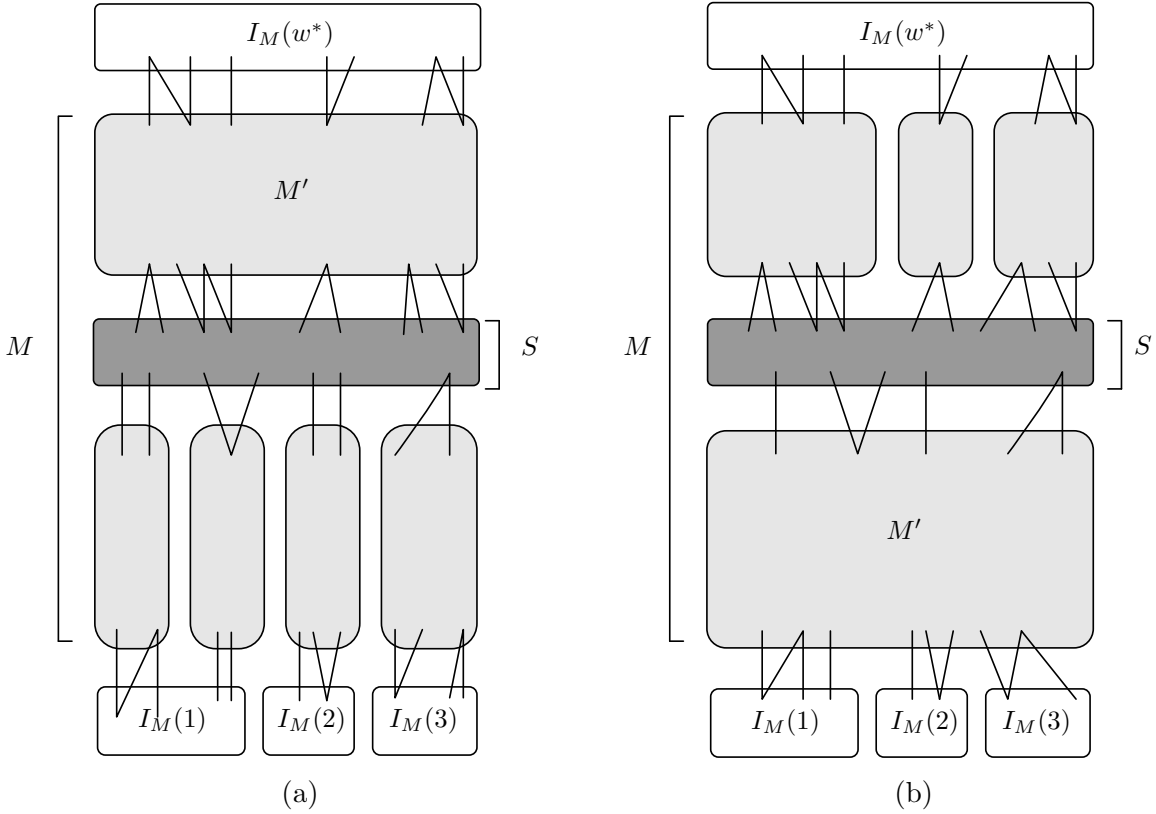

\begin{center}
{\small \svg{\linewidth}{goodmway}}
\caption{The two possibilities in Case 2 of Lemma~\ref{IterationStep}
  (the set of legs is $\wleg=\{w^*,1,2,3\}$): either (a)
  $N(M')\subseteq I_M(w^*)\cup S$ or (b) $N(M')\subseteq
  I_M(\wleg\setminus \{w^*\})\cup S$ holds.}\label{fig:goodmway}
\end{center}
\end{figure}

To show that $|I_{M'}(w^*)\setminus \wleg|$ and
$|I_{M'}(\wleg\setminus \{w^*\})\setminus \wleg|$ are both at most
$p$, we argue as follows.  Suppose first that $N(M')\subseteq
I_M(w^*)\cup S$. We show that $I_{M'}(w^*)\subseteq I_M(w^*)$ and
$I_{M'}(\wleg\setminus \{w^*\})\subseteq S$ hold, proving the bounds $|I_{M'}(w^*)\setminus \wleg|\le |I_M(w^*)\setminus \wleg|\le {\size}$ and $|I_{M'}(\wleg\setminus
\{w^*\})\setminus\wleg|\le |S|\le {\size}$.  Let $C_1$ (resp., $C_2$) be the union of
all those components of $G[\wleg\cup (K\setminus M')]$ that contain a
vertex of $w^*$ (resp., a vertex of $\wleg\setminus \{w^*\}$). As $M'$
is a multiway cut, $C_1$ and $C_2$ are disjoint. Now $I_{M'}(w^*)$ and
$I_{M'}(\wleg \setminus \{w^*\})$ are precisely the neighbors of $M'$
in $C_1$ and $C_2$, respectively. We observe that $I_M(w^*)\subseteq
C_1$: if $v\in I_M(w^*)$ is not in $C_1$, then every $w^*-v$ path has
to go through $M'\subset M$, contradicting the definition of
$I_M(w^*)$.  Thus $N(M')\subseteq I_M(w^*)\cup S$ implies that every
neighbor of $M'$ in $C_2$ is from $S$ (as it cannot be from
$I_M(w^*)\subseteq C_1$), further implying $I_{M'}(\wleg\setminus
\{w^*\})\subseteq S$.  Next, we show that $S\subseteq C_2$.  Suppose
that there is a $v\in S\setminus C_2$ and a $w^*-\wleg\setminus
\{w^*\}$ path $P$ intersecting $S$ only in $v$ (recall that $S$ is a
minimal $w^*-\wleg\setminus \{w^*\}$ separator). However, when the
path $P$ enters $C_2$ from $M'$, then, as we have seen, it enters a
vertex of $S\cap C_2$ that is different from $v$, a
contradiction. Thus $N(M')\subseteq I_M(w^*)\cup S$ implies that every
neighbor of $M'$ in $C_1$ is from $I_M(w^*)$ (as it cannot be from
$S\subseteq C_2$), further implying $I_{M'}(w^*)\subseteq I_M(w^*)$.
Finally, we can deduce that $N(M')=I_{M'}(\wleg)$, as required by the
definition of good multiway cut: indeed, every vertex of
$N(M')\subseteq I_M(w^*)\cup S$ is in $C_1\cup C_2$, that is, either
in $I_{M'}(w^*)$ or in $I_{M'}(\wleg\setminus\{w^*\})$. Therefore, we
have shown that $M'\subset M$ is a good multiway cut.

A symmetrical argument (exchanging the role of $w^*$ and
$\wleg\setminus \{w^*\}$) shows that if $N(M')\subseteq
I_M(\wleg\setminus \{w^*\})\cup S$, then $I_{M'}(w^*)\subseteq S$ and
$I_{M'}(\wleg \setminus \{w^*\})\subseteq I_M(\wleg \setminus
\{w^*\})$, implying the bounds $|I_{M'}(w^*)\setminus \wleg|\le
{\size}$ and $|I_{M'}(\wleg\setminus \{w^*\})\setminus \wleg|\le
{\size}$.  Thus in both cases, we proved that $M'\subset M$ is a good
multiway cut.

\textbf{Case 3.} Assume now that the algorithm returns $B:=(S \cup
N(M)) \setminus \wleg$ as a shattering set.  This happens because the number of components of $G[K \setminus (N(M)\cup S)]$ which are
multiway cuts of $(G[K\cup \wleg],\wleg)$ is not exactly one.  According to Lemma
\ref{GraceCriterion}, $N(M)\cup S$ is indeed a shattering set in this
case. Clearly, its size is at most $3p$.
\end{proof}

Lemma~\ref{lem:findshattering} follows by iterative
application of Lemma \ref{IterationStep}.
\begin{proof}[Proof (of Lemma~\ref{lem:findshattering})]

It is not hard to see that $K$ is a good multiway cut
of $(G[K \cup \wleg],\wleg)$; in particular,  $I(w)=\{w\}$ for every $w\in \wleg$, and hence $I(w^*)\setminus \wleg=I(\wleg \setminus \{w^*\})=\emptyset$. Let $M_0=K$. Apply the algorithm
of Lemma \ref{IterationStep} to $M_0$. The algorithm either
returns a shattering set of size at most $3p$ or a good multiway
cut $M_1 \subset M_0$. In the former case, we return the shattering set, 
in the latter case, apply the algorithm of Lemma \ref{IterationStep}
to $M_1$. Continuing this way, we obtain a sequence 
$M_0 \supset M_1 \supset \dots$ of good multiway cuts of decreasing
size. It follows that after at most $|V(G)|$ iterative applications of the
algorithm of Lemma \ref{IterationStep}, a shattering set of size at
most $3p$ will be returned. \end{proof}

\section{Finding a shadowless solution by reduction to
  Almost 2SAT}
\label{sec:find-nonis-solut}

The goal of this section is to show that we can solve \bmcc\ if we
know that there is at least one shadowless solution.

Let $x_1$, $\dots$, $x_n$ be a set of variables; a literal is either a
variable $x_i$ or its negation $\overline x_i$.  Recall that a 2CNF
formula is a conjunction of clauses with at most two literals in each
clause, e.g., $(\overline x_1 \vee x_2) \wedge (\overline x_3) \wedge
(x_1 \vee \overline x_4 )$. The classical 2SAT problem asks if a given
2CNF formula has a satisfying assignment. It is well-known that a
satisfying assignment for a 2CNF formula can be found in linear time
(if exists). However, it is NP-hard to find an assignment that
maximizes the number of satisfied clauses, or equivalently, to find a
minimum set of clauses whose removal makes the formula
satisfiable. Lokshtanov et al.~\cite{DBLP:journals/corr/abs-1203-0833}
(improving earlier work
\cite{DBLP:journals/jcss/RazgonO09,DBLP:journals/toct/CyganPPW13,DBLP:conf/esa/RamanRS11}) gave an
$O^*(2.3146^k)$ time algorithm for the problem of deciding if a 2CNF
formula can be made satisfiable by the deletion of at most $k$
clauses; they call this problem \textsc{Almost 2SAT}. We need a
variant of the result here, where instead of deleting at most $k$
clauses, we are allowed to delete at most $k$ variables.  An easy
reduction (see Appendix~\ref{app:almost2sat}) gives an algorithm
for this variant.  If $\phi$ is a 2CNF formula and $X$ is a set of
variables, then we denote by $\phi\setminus X$ the formula obtained by
removing every clause containing a literal of a variable in $X$.

\begin{theorem}\label{th:almost2satx}
  Given a \textup{2CNF} formula $\phi$ and an integer $k$, in time $O^*(2.3146^k)$ we can either find a set $X$ of
  at most $k$ variables such that $\phi\setminus X$ is satisfiable, or
  correctly state that no such set $X$ exists.
\end{theorem}

It is not difficult to reduce finding a shadowless solution to the
problem solved by Theorem~\ref{th:almost2satx}. For each vertex $v$ of
$G\setminus W$, we introduce a variable whose value expresses which
leg of the component containing $v$ is reachable from $v$. This
formulation cannot express that a vertex is separated from both legs.
However, as we assume that there is a shadowless solution, we do not
have to worry about such vertices.

\begin{proof}[Proof (of Lemma~\ref{lem:shadowless2sat})]
  We encode the \bmcc\ instance $I=(G,{\bf T},W,{\size})$ as a 2CNF formula $\phi$ the
  following way. For each component $C$ of $G\setminus W$ having two
  legs, let $\ell_0(C)$ and $\ell_1(C)$ be the two legs. If component
  $C$ has only one leg, then let $\ell_0(C)$ be this leg, and let
  $\ell_1(C)$ be undefined.  For every vertex $v\in C$, let
  $\ell_0(v)=\ell_0(C)$ and $\ell_1(v)=\ell_1(C)$. We construct a
  formula $\phi$ whose variables correspond to $V(G)\setminus W$. The
  intended meaning of the variables is that $v$ has value
  $b\in\{0,1\}$ if 
  $v$ is in the same component as $\ell_b(v)$ after removing the
  solution. To enforce this interpretation, $\phi$ contains the
  following clauses:
\begin{itemize}
\item Group 1: $(u\to v)$, $(v\to u)$ for every adjacent $u,v\in V(G)\setminus
  W$.
\item Group 2: If $u$ is a neighbor of $\ell_b(u)$ for some $b\in\{0,1\}$, then
  there is a clause $(u=b)$.
\item Group 3: If $(u,v)\in {\bf T}$, $u,v\not\in W$, and $\ell_{b_u}(u)=\ell_{b_v}(v)$ for some
  $b_u,b_v\in \{0,1\}$, then there is a clause $(u\neq b_u \vee v\neq
  b_v)$ (e.g., if $\ell_0(u)=\ell_1(v)$, then the clause is $(u\vee
  \overline v)$).
\item Group 4: If $(u,v)\in {\bf T}$, $u\in W$, $v\not \in W$, and
  $\ell_b(v)=u$ for some $b\in \{0,1\}$, then there is a clause
  $(v\neq b)$.
\end{itemize}
This completes the description of $\phi$. Note that no clause is
introduced for pairs $(u,v)\in {\bf T}$ with $u,v\in W$, but these
pairs are automatically separated by a solution that is a multiway cut
of $W$. Furthermore, we can assume that $W$ induces an independent set,
otherwise there is no solution.

We show first that if $I$
has a shadowless solution $S$, then removing the corresponding
variables of $\phi$ makes it satisfiable. As $S$ is shadowless and
it is a multiway cut of $W$, every vertex of $G\setminus S$ is in the same
component as exactly one of $\ell_0(v)$ and $\ell_1(v)$; let the value
of variable $v$ be $b$ if vertex $v$ is in the same component as
$\ell_b(v)$. It is clear that this assignment satisfies the clauses in
the first two groups. Consider a clause $(u\neq b_u \vee v\neq b_v)$
from the third group. This means that $(u,v)\in {\bf T}$ and
$\ell_{b_u}(u)=\ell_{b_v}(v)=w\in W$. If this clause is not satisfied,
then $u=b_u$ and $v=b_v$. By the way the assignment was defined, this
is only possible if $u$ is in the same component of $G\setminus S$ as
$\ell_{b_u}(u)=w$ and $v$ is in the same component of $G\setminus S$
as $\ell_{b_v}(v)=w$. Therefore, $u$ and $v$ are in the same component
of $G\setminus S$, contradicting the assumption that $S$ is a solution
of $I$. Clauses in Group 4 can be checked similarly.

We have shown that $\phi$ can be made satisfiable by the deletion of
${\size}$ variables. By Theorem~\ref{th:almost2satx}, we can find such
a set $S'$ of variables in time $O^*(4^{\size})$. To complete the proof,
we show that such a set $S'$ corresponds to a (not necessarily
shadowless) solution of $I$.  Let us show first that $S'$ is a
multiway cut of $W$. Suppose that there is a path $P$ connecting
$w_0,w_1\in W$ in $G\setminus S'$. We can assume that the internal
vertices of $P$ are disjoint from $W$, i.e., they are in one component $C$ of
$G\setminus W$ with two legs. Thus there is a path $P'$ from a
neighbor $v_0$ of $w_0$ to a neighbor $v_1$ of $w_1$ in $C\setminus
S'$.  Suppose without loss of generality that $\ell_0(C)=w_0$ and
$\ell_1(C)=w_1$. As the clauses in Group 1 are satisfied, every
variable of $P'$ has the same value. However, because of the clauses
in Group 2, we have $x_{v_0}=0$ and $x_{v_1}=1$, a contradiction. Therefore,
we can assume that $S'$ is a multiway cut of $W$.

Suppose now that there is some $(u,v)\in {\bf T}$ such that $u,v\not
\in W$ are in the same component of $G\setminus S'$; let $P$ be a
$u-v$ path in $G\setminus S'$. As $W$ is a multicut of ${\bf T}$, it
is clear that $P$ goes through at least one vertex of $W$. We have
seen that $S'$ is a multiway cut of $W$, thus $P$ goes through exactly one
vertex of $W$. Let $P=P_1 w P_2$ for some path $P_1$ that is fully
contained in the component of $G\setminus W$ containing $u$ and path
$P_2$ fully contained in the component containing $v$. Let $b_u,b_v\in
\{0,1\}$ be such that $\ell_{b_u}(u)=\ell_{b_v}(v)=w$. Group 1 ensures
that every variable of $P_1$ has the same value and Group 2 ensures
that the last variable of $P_1$ has value $b_u$, thus $u=b_u$. A
similar argument shows that $v=b_v$. However, this means that clause
$(u\neq b_u \vee v\neq v_u)$ of Group 3 is not satisfied, a
contradiction. Finally, a similar argument shows that the clauses in
Group 4 ensure that pairs $(u,v)\in {\bf T}$ with $u\in W$, $v\not\in
W$ are separated.
\end{proof}

\section{Hardness of {\rm \textsc{Directed Multicut}}}
\label{sec:hardn-rm-textscd}
We prove that \textsc{Directed Edge Multicut} is W[1]-hard
parameterized by the solution size, thus it is not fixed-parameter
tractable (assuming the widely-held complexity hypothesis $\textup{FPT}\neq
\textup{W[1]}$). Recall that the edge and vertex versions are
equivalent, thus the hardness result holds for both versions. The proof
below proves the hardness result for the weighted version of the
problem, where each edge has a positive integer weight, and the task
is to find a multicut with total weight at most ${\size}$. If the
weights are polynomial in the size of the input (which is true in the
proof), then the weighted version can be reduced to the unweighted
version by introducing parallel edges. Thus the proof proves the
hardness of the unweighted version as well. For notational
convenience, we allow edges with weight $\infty$; such edges can be
easily replaced by edges with sufficiently large finite weight.

\begin{theorem}\label{th:dirmulti}
\textsc{Directed Edge Multicut} is \textup{W[1]}-hard parameterized by the size
${\size}$ of the cutset.
\end{theorem}
\begin{proof}
  We prove hardness for the weighted version of the problem by parameterized
  reduction from \textsc{Clique}. Let $G$ be a graph with $m$ edges
  and $n$ vertices where a clique of size $t$ has to be found. We
  construct an instance of \textsc{Directed Edge Multicut} containing
  $t(t-1)$ gadgets: for each ordered pair $(i,j)$ ($1\le i,j\le t$,
  $i\neq j$), there is a gadget $G_{i,j}$. Intuitively, each gadget
  $G_{i,j}$ has $2m$ possible states and a state represents an ordered
  pair $(v_i,v_j)$ of adjacent vertices. We would like to ensure that the gadgets
  describe a $t$-clique $\{v_1,\dots,v_t\}$ in the sense that $G_{i,j}$ represents the pair
  $(v_i,v_j)$. In order to enforce this
  interpretation, we need to connect the gadgets in a way that
  enforces two properties:
\begin{itemize}
\item[(1)] if $G_{i,j}$
  represents $(v_i,v_j)$, then $G_{j,i}$ represents $(v_j,v_i)$, and
\item[(2)] if $G_{i,j}$ represents $(v_{i},v_{j})$ and $G_{i,j'}$
  represents $(u_i,u_j)$, then $v_i=u_i$. 
\end{itemize}
(Note that  it follows from these two conditions that if $G_{i,j}$
and $G_{i',j}$ represent $(v_i,v_j)$ and $(u_i,u_j)$, respectively,
then $v_j=u_j$.)

Let us identify the vertices of $G$ with the integers $0$, $\dots$,
$n-1$ and let us define $\iota(x,y)=xn+y$, which is a bijective
mapping from $\{0,\dots,n-1\}\times \{0,\dots,n-1\}$ to
$\{0,\dots,n^2-1\}$. The gadget $G_{i,j}$ has $n^2+1$ vertices
$w_{i,j}^0$, $\dots$, $w_{i,j}^{n^2}$. Let $D:=2t^2$. For every $0 \le
s <n^2$, there is an edge $\ora{w_{i,j}^sw_{i,j}^{s+1}}$ whose weight
is $D$ if $\iota^{-1}(s)$ is a pair $(x,y)$ such that $x$ and $y$ are
adjacent in $G$, and $\infty$ otherwise. Furthermore, there is an
additional edge $\ora{w_{i,j}^{n^2}w_{i,j}^0}$ with weight $\infty$.
The \textsc{Directed Edge Multicut} instance contains the terminal
pair $(w_{i,j}^0,w_{i,j}^{n^2})$, which means that at least one of the
edges $\ora{w_{i,j}^s,w_{i,j}^{s+1}}$ having finite weight has to be
in the multicut. If the multicut contains exactly one such edge
$\ora{w_{i,j}^s,w_{i,j}^{s+1}}$ in the gadget, then we say that the
gadget {\em represents} the pair $\iota^{-1}(s)$.  We set
${\size}:=t(t-1)D+t+t(t-1)/2$ to be the maximum weight of the multicut. Since
${\size}<t(t-1)D+D$, a multicut of weight at most ${\size}$ contains exactly one
edge of weight $D$ from each gadget, implying that each gadget
represents some pair.

\begin{figure}[t]
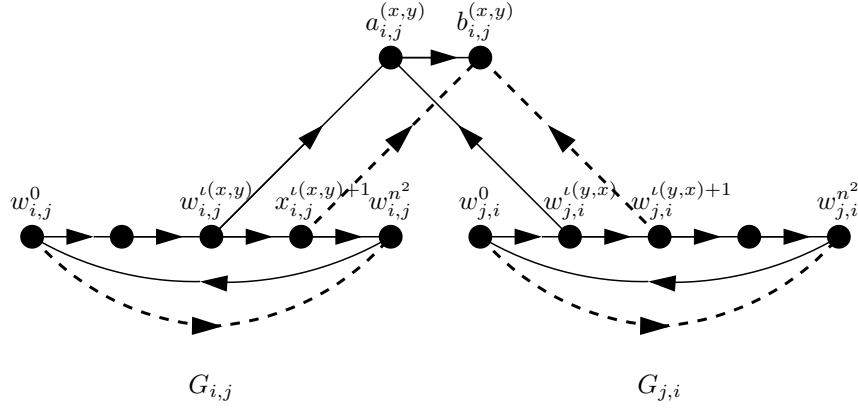

\begin{center}
{\small \svg{0.7\linewidth}{cycle}}
\caption{Part of a connection between gadgets $G_{i,j}$ and $G_{j,i}$
  in the proof of Theorem~\ref{th:dirmulti}. The dashed edges
  represent terminal pairs.
}\label{fig:dirhard}
\end{center}
\end{figure}

  For every $1\le i < j \le t$, we connect $G_{i,j}$ and $G_{j,i}$ in
  a way that ensures that if $G_{i,j}$ represents the pair $(x,y)$, then
  $G_{j,i}$ represents the pair $(y,x)$ (see Fig.~\ref{fig:dirhard}). More precisely, we show that
  if the multicut contains exactly one edge of the connection and
  $G_{i,j}$ (resp., $G_{j,i}$) represents the pair $(x,y)$ (resp.,
  $(x',y')$), then $x=y'$ and $y=x'$. For every ordered pair $(x,y)$
  of adjacent vertices of $G$, let us introduce two new vertices $a_{i,j}^{(x,y)}$,
  $b_{i,j}^{(x,y)}$, and the directed edge
  $\ora{a_{i,j}^{(x,y)},b_{i,j}^{(x,y)}}$ having weight $1$.
  Furthermore, let us add the edges
  $\ora{w_{i,j}^{\iota(x,y)}a_{i,j}^{(x,y)}}$ and
  $\ora{w_{j,i}^{\iota(y,x)}a_{i,j}^{(x,y)}}$ having weight $\infty$. Finally, let us add the
  terminal pairs $(w_{i,j}^{\iota(x,y)+1},b_{i,j}^{(x,y)})$ and
  $(w_{j,i}^{\iota(y,x)+1},b_{i,j}^{(x,y)})$. Observe that if
  $G_{i,j}$ represents $(x,y)$, then $w_{i,j}^{\iota(x,y)}$ (and hence
  $a_{i,j}^{(x,y)}$) is reachable from $w_{i,j}^{\iota(x,y)+1}$, which
  means that the multicut has to contain the edge
  $\ora{a_{i,j}^{(x,y)},b_{i,j}^{(x,y)}}$. Similarly, if $G_{j,i}$
  represents $(x',y')$, then $w_{j,i}^{\iota(x',y')}$ (and hence
  $a_{i,j}^{(y',x')}$) is reachable from $w_{j,i}^{\iota(x',y')+1}$,
  which means that the multicut has to contain the edge
  $\ora{a_{i,j}^{(y',x')},b_{i,j}^{(y',x')}}$. If the multicut
  contains only one edge of the connection between $G_{i,j}$ and
  $G_{j,i}$, the two edges must coincide, and we have $x=y'$, $y=x'$.

  For every $1\le i \le t$ and $0 \le x < n$, we introduce two new
  vertices $c_{i}^x$ and $d_{i}^x$ and connect them with the edge
  $\ora{c_i^x d_{i}^x}$ having weight 1. For every $1 \le j \le t$,
  $i\neq j$, $0\le x <n$, we add an edge
  $\ora{w_{i,j}^{\iota(x,0)}c_i^x}$ having weight $\infty$ and a
  terminal pair $(w_{i,j}^{\iota(x+1,0)},d_i^x)$. This completes the
  description of the reduction. Note that if $G_{i,j}$
  represents $(x,y)$, then $\iota(x,0) \le \iota(x,y) <\iota(x+1,0)$
  implies that $w_{i,j}^{\iota(x,0)}$ is reachable from
  $w_{i,j}^{\iota(x+1,0)}$, which means that the edge
  $\ora{c_i^xd_i^x}$ has to be in the cut to prevent $d_i^x$ from
  being reachable from $x_{i,j}^{\iota(x+1,0)}$.

  Suppose there is a multicut of
  weight at most ${\size}$.  This means that the multicut contains at most
  $t(t-1)$ edges of weight $D$, thus each gadget $G_{i,j}$ contains
  exactly one edge of weight $D$, i.e., each gadget represents some
  pair $(x,y)$. As discussed in the previous two paragraphs, if
  $G_{i,j}$ represents $(x,y)$, then $\ora{c_{i}^xd_i^x}$ is in the
  multicut.  Furthermore, depending on whether $i<j$ or $i>j$ holds,
  either $\ora{a_{i,j}^{(x,y)}b_{i,j}^{(x,y)}}$ or
  $\ora{a_{j,i}^{(y,x)}b_{j,i}^{(y,x)}}$, respectively,  is in the multicut as well.
  If the weight of the multicut is at most ${\size}$, then the total weight
  of these edges is at most $t+t(t-1)$, which is only possible if
  these edge coincide in every possible way and it follows that
  properties (1) and (2) hold. Therefore, there are distinct vertices
  $v_{1}$, $\dots$, $v_t$ such that gadget $G_{i,j}$ represents
  $(v_i,v_j)$, which implies that $v_1$, $\dots$, $v_t$ is a clique in
  $G$.

For the other direction, suppose that $v_1$, $\dots$, $v_t$ is a
clique in $G$. Let us consider the multicut that contains the
following edges:
\begin{itemize}
\item $\ora{w_{i,j}^{\iota(v_i,v_j)}w_{i,j}^{\iota(v_i,v_j)+1}}$ for
  every $1 \le i,j \le t$, $i\neq j$,
\item $\ora{a_{i,j}^{(v_i,v_j)}b_{i,j}^{(v_i,v_j)}}$ for every $1 \le i
  <j\le t$, and
\item $\ora{c_i^{v_i}d_i^{v_i}}$ for every $1\le i \le t$.
\end{itemize}
The total weight of these edges is exactly ${\size}$. 
The edges in the first group ensure that 
$w_{i,j}^{n^2}$ is not reachable from $w_{i,j}^0$ for any $i,j$.
For some $i<j$ and adjacent vertices $x$ and $y$, consider a terminal pair
$(w_{i,j}^{\iota(x,y)+1},b_{i,j}^{\iota(x,y)})$. If $(x,y)\neq
(v_i,v_j)$, then edge
$\ora{w_{i,j}^{\iota(v_i,v_j)}w_{i,j}^{\iota(v_i,v_j)+1}}$ of the
multicut ensures that 
$w_{i,j}^{\iota(x,y)}$ (and hence $b_{i,j}^{\iota(x,y)}$) is not
reachable from $w_{i,j}^{\iota(x,y)+1}$. If $(x,y)=(v_i,v_j)$, then
edge $\ora{a_{i,j}^{(v_i,v_j)}b_{i,j}^{(v_i,v_j)}}$ is in the multicut,
again disconnecting this terminal pair. For $i>j$, an analogous
argument shows that  terminal pair
$(w_{i,j}^{\iota(x,y)+1},b_{j,i}^{\iota(y,x)})$ for every $x,y$ is
disconnected. Consider now the terminal pair
$(w_{i,j}^{\iota(x+1,0)},d_i^x)$ for some $1 \le i,j \le t$, $i\neq
j$, $0 \le x <n$. If $x\neq v_i$, then $\iota(v_i,v_j)$ is either less
than $\iota(x,0)$ or at least $\iota(x+1,0)$, thus the edge
$\ora{w_{i,j}^{\iota(v_i,v_j)}w_{i,j}^{\iota(v_i,v_j)+1}}$ of the
multicut 
ensures that $w_{i,j}^{\iota(x,0)}$ (and hence $d_{i}^x$) is not
reachable from $w_{i,j}^{\iota(x+1,0)}$. On the other hand, if
$x=v_i$, then the edge $\ora{c_i^{v_i}d_i^{v_i}}$ is in the third
group of the multicut. Thus we have shown that if there is a clique of
size $t$ in $G$, then there is a multicut of size at most ${\size}$.
\end{proof}
\section*{Acknowledgement}
We would like to thank the reviewers for the insightful comments
that have helped us to fix a number of errors and improve readability. 



\appendix

\section{Important separators}
\label{app:important-separators}

First we state without proof some properties of Definition~\ref{def:important} that are easy to see:
\begin{proposition}\label{prop:importantprop}
Let $G$ be a graph, $X,Y\subseteq V(G)$ be two disjoint sets of
vertices, and $S$ be an important $X-Y$ separator.
\begin{enumerate}
\item\label{prop:item:del} For every $v\in S$, the set $S\setminus \{v\}$ is an important $X-Y$
  separator in $G\setminus v$.
\item \label{prop:item:increase} If $S$ is an $X'-Y$ separator for some $X'\supset X$, then $S$
  is an important $X'-Y$ separator.
\end{enumerate}
\end{proposition}

\begin{proof}[Proof (of Lemma~\ref{lem:impsep})]
 We prove
  by induction on $2{\size}-\lambda$ that there are at most $2^{2{\size}-\lambda}$ important $X-Y$
  separators of size at most ${\size}$, where 
  $\lambda$ is the size of the smallest $X-Y$ separator.
 If $\lambda>{\size}$, then there is no
  $X-Y$ separator of size ${\size}$, and therefore the statement holds if
  $2{\size}-\lambda<0$. Also, if $\lambda=0$ and ${\size}\ge 0$, then there is a
  unique important $X-Y$ separator of size at most ${\size}$: the empty set.

  If $S$ is an $X-Y$ separator, then we denote by $K_S$ the union of every
  component of $G\setminus S$ intersecting $X$. First we show the
  well-known fact that there is a unique $X-Y$ separator $S^*$ of size
  $\lambda$ such that $K_{S^*}$ is inclusionwise maximal, i.e., we
  have $K_{S}\subset K_{S^*}$ for every other $X-Y$ separator $S$ of
  size $\lambda$.  Suppose that there are two separators $S'$ and
  $S''$ such that $K_{S'}$ and $K_{S''}$ are incomparable and
  inclusionwise maximal. Let us define the function
  $\gamma(Z)=|N(Z)|$. It is well-known that $\gamma$ is submodular, that is,
\[
\gamma(A)+\gamma(B)\ge \gamma(A\cup B)+\gamma(A\cap B)
\]
for every $A,B\subseteq V(G)$.  In particular, the submodularity of
gamma implies that
\begin{align*}
  \underbrace{\gamma(K_{S'})}_{=\lambda}+\underbrace{\gamma(K_{S''})}_{=\lambda}\ge
\gamma(K_{S'}\cup K_{S''})+
\underbrace{\gamma(K_{S'}\cap K_{S''})}_{\ge \lambda}.
\end{align*}
The left hand side is exactly $2\lambda$, while the second term of the
right hand side is at least $\lambda$ (as $N(K_{S'}\cap
K_{S''})$ is an $X-Y$ separator). Therefore,
$\gamma(K_{S'} \cup K_{S''})\le \lambda$. This means that
$N(K_{S'}\cup K_{S''})$ is also a minimum $X-Y$ separator, contradicting the
maximality of $S'$ and $S''$.

Next we show that for every important $X-Y$ separator $S$, we have
$K_{S^*}\subseteq K_S$. Suppose this is not true for some $S$. We use
submodularity again:
\[
\underbrace{\gamma(K_{S^*})}_{=\lambda}+\gamma(K_{S})\ge
\gamma(K_{S^*}\cup K_{S})+
\underbrace{\gamma(K_{S^*}\cap K_{S})}_{\ge \lambda}.
\]
By definition, $\gamma(K_{S^*})=\lambda$, and $N(K_{S^*}\cap K_{S})$
is an $X-Y$ separator, hence $\gamma(K_{S^*}\cap K_{S})\ge
\lambda$. This means that $\gamma(K_{S^*}\cup K_{S})\le
\gamma(K_{S})$. However this contradicts the assumption that $S$ is an
important $X-Y$ separator: $N(K_{S^*}\cup K_{S})$ is an $X-Y$
separator not larger than $S$, but $K_{S^*}\cup K_{S}$ is a proper
superset of $K_S$ (as $K_{S^*}$ is not a subset of $K_S$ by
assumption).

We have shown that for every important separator $S$, the set $K_S$
contains $K_{S^*}$. Let $v\in S^*$ be an arbitrary vertex of $S^*$
(note that $\lambda>0$, hence $S^*$ is not empty). An important $X-Y$
separator $S$ of size at most ${\size}$ either contains $v$ or not. If $S$
contains $v$, then $S\setminus \{ v\}$ is an important $X-Y$ separator in $G\setminus
v$ of size at most ${\size}':={\size}-1$
(Prop.~\ref{prop:importantprop}(\ref{prop:item:del})). As $v\not\in X,Y$, the size $\lambda'$ of the
minimum $X-Y$ separator in $G\setminus v$ is at least $\lambda-1$.
Therefore, $2{\size}'-\lambda'<2{\size}-\lambda$ and the induction hypothesis
implies that there are at most $2^{2{\size}'-\lambda'}\le 2^{2{\size}-\lambda-1}$
important $X-Y$ separators of size ${\size}'$ in $G\setminus \{v\}$, and
hence at most that many important $X-Y$ separators of size ${\size}$ in $G$
that contain $v$.

Let us count now the important $X-Y$ separators not containing $v$.
Note that by the minimality of $S^*$, vertex $v$ of $S^*$ has a neighbor
in $K_{S^*}$. We have seen that $K_{S^*}\subseteq K_S$ for every such
$X-Y$ separator $S$. As $v\not \in S$ and $v$ has a neighbor in $K_S$,
even $K_{S^*}\cup\{v\}\subseteq K_S$ is true. Let
$X'=K_{S^*}\cup\{v\}$; it follows that $S$ is a $X'-Y$ separator and
in fact an important $X'-Y$ separator by Prop.~\ref{prop:importantprop}(\ref{prop:item:increase}). 
There is no $X'-Y$ separator $S$ of size $\lambda$: such a set $S$ would be an
$X-Y$ separator of size $\lambda$ as well, with $K_{S^*}\cup
\{v\}\subseteq K_S$, contradicting the maximality of $S^*$. Thus the
minimum size $\lambda'$ of an $X'-Y$ separator is greater than
$\lambda$. It follows by the induction assumption that the number of
important $X'-Y$ separators of size at most ${\size}$ is at most
$2^{2{\size}-\lambda'}\le 2^{2{\size}-\lambda-1}$, which is a bound on the number
of important $X-Y$ separators of size ${\size}$ in $G$ that does not
contain $v$.

Adding the bounds in the two cases, we get the required bound
$2^{2{\size}-\lambda}$. An algorithm for enumerating all the at most $4^{\size}$
important separators follows from the above proof. First, we can find
a maximum $X-Y$ flow in time $O({\size}(|V(G)|+|E(G)|))$ using at most ${\size}$
rounds of the Ford-Fulkerson algorithm. It is well-known that the
separator $S^*$ in the proof  can be
deduced from the maximum flow in linear time by finding those vertices
from which $Y$ cannot be reached in the residual graph
\cite{MR0159700}.  Pick any arbitrary vertex $v\in S^*$. Then we
branch on whether vertex $v\in S^{*}$ is in the important separator or
not, and recursively find all possible important separators for both
cases.    Note that this algorithm
enumerates a superset of all important separators: by our analysis
above, every important separator is found,
but there is no guarantee that all the constructed separators are
important. Therefore, the algorithm has to be followed by a filtering
phase where we check for each returned separator whether it is
important. Observe that $S$ is an important $X-Y$ separator if and
only if $S$ is the unique minimum $K_S-Y$
separator, where $K_S$ is the set of vertices reachable from $X$ in $G\setminus S$. As the size of $S$ is at most ${\size}$, this can be checked in
time $O({\size}(|V(G)|+|E(G)|))$ by finding a maximum flow and constructing
the residual graph.  The search tree has at most $4^{\size}$ leaves and the
work to be done in each node is $O({\size}(|V(G)|+|E(G)|))$. Therefore, the
total running time of the branching algorithms is $O(4^{\size}\cdot
{\size}(|V(G)|+|E(G)|))$ and returns at most $4^{\size}$ separators. This is followed by the
filtering phase, which takes time $O(4^{\size}\cdot {\size}(|V(G)|+|E(G)|))$. 
\end{proof}

\section{Deleting variables in \textsc{Almost 2SAT}}
\label{app:almost2sat}

\begin{proof}[Proof (of Theorem~\ref{th:almost2satx})]
  Let $x_1$, $\dots$, $x_n$ be the variables of $\phi$. We create a
  new 2CNF formula $\phi'$ on $2n$ variables $x^b_i$ ($1\le i \le n$,
  $b\in \{0,1\}$). The intended meaning of $x^b_i$ is that its value
  is 1 if and only if the value of $x_i$ in $\phi$ is $b$. For every
  $1\le i \le n$, let us introduce a clause $(\overline x^0_i \vee \overline
  x^1_i)$ in formula $\phi'$. For every clause of $\phi$, there is a
  corresponding clause of $\phi'$ where literal $x_i$ is replaced by
  literal $x^1_i$ and literal $\overline x_i$ is replaced by $x^0_i$ (e.g.,
  $(x_i \vee \overline x_j)$ is replaced by $(x^1_i \vee x^0_j)$.

We claim that there is a set $X$ of variables in $\phi$ such that
$\phi\setminus X$ is satisfiable if and only if there is a set $X'$ 
($|X|=|X'|$) of clauses in $\phi'$ such that $\phi'\setminus X'$ is satisfiable. As
the existence of such a $X'$ can be tested by the algorithm of
\cite{DBLP:journals/jcss/RazgonO09} in time $O^*(4^k)$, the theorem
follows from this claim.

Suppose first that there is such a set $X$ of variables in $\phi$; let
$f$ be a satisfying assignment of $\phi\setminus X$. Let $X'$ contain
the clauses $(\overline x^0_i\vee \overline x^1_i)$ for every $x_i\in X$. Let us
define $f'(x^0_i)=f'(x^1_i)=1$ if $x_i\in X$, and for every
$x_i\not\in X$, let $f'(x^b_i)=1$ if
and only if $f(x_i)=b$. It is straightforward to verify that $f'$
satisfies $\phi'\setminus X'$.

For the other direction, let us suppose that $X'$ is a set of clauses
such that $\phi'\setminus X'$ is satisfiable and let $f'$ be a
satisfying assignment of $\phi'\setminus X'$. The important
observation is that we can assume that $X'$ contains only clauses of
the form $(\overline x^0_i\vee \overline x^1_i)$.  To see this, observe that variables
$x^b_i$ appear negatively only in the clauses of this form. Thus if
$X'$ contains a clause $C$ that where $x^b_i$ appears positively, then
we can replace $C$ in $X'$ by $(\overline x^0_i\vee \overline x^1_i)$ and set
$f(x^0_i)=f(x^1_i)=1$. Let $X$ contain a variable $x_i$ if $(\overline
x^0_i\vee \overline x^1_i)$ is in $X'$. It is easy to verify that defining
$f(x_i)=f(x^1_i)$ gives a satisfying assignment of $\phi\setminus X$.
\end{proof}


\end{document}